\documentclass[preprint,numbers]{elsarticle}

\ifdefined\pdfminorversion\pdfminorversion=7\fi

\usepackage{graphicx}%
\usepackage{multirow}%
\usepackage{amsmath,amssymb,amsfonts}%
\usepackage{mathrsfs}%
\usepackage{xcolor}%
\usepackage{textcomp}%
\usepackage{booktabs}%
\usepackage{colortbl}
\usepackage[colorlinks=true,linkcolor=blue,citecolor=blue,urlcolor=blue,breaklinks=true]{hyperref}
\usepackage[nameinlink]{cleveref}

\newcommand{\sref}[1]{\cref{#1}}

\crefname{figure}{Fig.}{Figs.}
\Crefname{figure}{Figure}{Figures}
\crefname{extfigure}{Fig.}{Figs.}
\Crefname{extfigure}{Fig.}{Figs.}
\crefname{table}{Table}{Tables}
\Crefname{table}{Table}{Tables}
\crefname{exttable}{Table}{Tables}
\Crefname{exttable}{Table}{Tables}

\begin{document}

\begin{frontmatter}

\title{Data-efficient crack quantification in lithium-ion cathodes using foundation model transfer}

\author[1,2,3]{Thorsten Tegetmeyer-Kleine\corref{cor1}}
\ead{thorsten.tegetmeyer-kleine@isea.rwth-aachen.de}

\author[7,8]{Thomas Schmitt}

\author[6]{Phillip Aquino}

\author[2,3,4]{Christiane Rahe}

\author[2,3,4,5]{Dirk Uwe Sauer}

\author[1,2,3]{Weihan Li\corref{cor1}}
\ead{weihan.li@isea.rwth-aachen.de}

\affiliation[1]{organization={Junior Professorship for Artificial Intelligence and Digitalization for Batteries, Institute for Power Electronics and Electrical Drives (ISEA), RWTH Aachen University},
            addressline={Campus-Boulevard 89},
            city={Aachen},
            postcode={52074},
            country={Germany}}

\affiliation[2]{organization={Center for Aging, Reliability and Lifetime Prediction of Electrochemical and Power Electronic Systems (CARL), RWTH Aachen University},
            addressline={Campus-Boulevard 89},
            city={Aachen},
            postcode={52074},
            country={Germany}}

\affiliation[3]{organization={Juelich Aachen Research Alliance, JARA-Energy},
            addressline={Templergraben 55},
            city={Aachen},
            postcode={52056},
            country={Germany}}

\affiliation[4]{organization={Chair for Electrochemical Energy Conversion and Storage Systems, Institute for Power Electronics and Electrical Drives (ISEA), RWTH Aachen University},
            addressline={Campus-Boulevard 89},
            city={Aachen},
            postcode={52074},
            country={Germany}}

\affiliation[5]{organization={Helmholtz Institute Muenster (HI MS), IMD-4, Forschungszentrum Juelich},
            country={Germany}}

\affiliation[6]{organization={Honda Research Institute USA, Inc.},
            addressline={21001 State Route 739},
            city={Raymond},
            state={OH},
            postcode={43067},
            country={United States}}

\affiliation[7]{organization={Honda Research Institute Europe GmbH},
            addressline={Carl-Legien-Strasse 30},
            city={Offenbach/Main},
            postcode={63073},
            country={Germany}}

\affiliation[8]{organization={Zenules GmbH},
            addressline={Landgraf-Georg-Str. 4},
            city={Darmstadt},
            postcode={64283},
            country={Germany}}

\cortext[cor1]{Corresponding authors.}

\begin{abstract}
Battery lifetime is central to sustainable electrification, yet the particle cracking that drives lithium-ion cathode aging is hard to measure: quantitative microscopy of this degradation is bottlenecked by annotation, because each destructive electron-microscopy cross-section spans hundreds of megapixels and pixel-level expert labelling requires hours per image. We show that a frozen self-supervised vision-transformer encoder, combined with a lightweight trainable decoder and iterative model-assisted annotation, turns this sparse labelling budget into population-scale degradation measurements. Applied to three 120-megapixel NMC cathode cross-sections representing initial, cycled-aged and calendar-aged states, the framework distinguishes intragranular cracks from early- and late-stage intergranular cracks and yields per-particle distributions of crack width, tortuosity and area fraction. Late intergranular crack coverage reaches 4.6\% in the cycled sample versus 0.5\% in the initial and calendar-aged samples, forming more tortuous, higher-coverage networks, consistent with degradation from repeated electrochemical cycling rather than elevated-temperature storage alone. A single destructive image yields the population-level statistics needed for lifetime-extending design, aging assessment and second-life decisions.
\end{abstract}


\end{frontmatter}

\section{Introduction}\label{sec:main}

Lithium-ion batteries are central to electrified transport and renewable-energy integration, yet electrochemical and chemo-mechanical degradation processes within their electrodes ultimately limit service life~\cite{Pender2020,Edge2021}. Quantifying the mechanical cracking that accompanies this degradation is bottlenecked by destructive cross-sectional microscopy and labour-intensive expert annotation. In high-nickel layered oxide NMC cathodes, this degradation appears primarily as particle-level cracking: secondary particles composed of aggregated primary grains develop cracks that propagate along grain boundaries (intergranular) and through individual grains (intragranular) during cycling, calendar aging, and thermal stress~\cite{Yan2017}. These cracks expose fresh cathode surfaces to the electrolyte, catalyzing parasitic reactions, promoting cathode-electrolyte interphase growth, and accelerating transition-metal dissolution, collectively driving capacity fade and impedance rise~\cite{heenan2020,OKane2021}. As cracking intensifies, secondary particles fragment entirely, causing loss of electronic contact and further capacity loss~\cite{Navidi2024,Britala2023}. Understanding crack initiation, propagation, and morphology is therefore central to diagnosing battery aging and guiding electrode design.

To quantify cracking reliably, one must first distinguish the underlying crack morphologies. Intergranular cracks propagate along grain boundaries between primary crystallites within polycrystalline secondary particles, yielding a fragmented, mosaic-like appearance, whereas intragranular cracks traverse individual grains and often align with the layered structure, running parallel to the (003) crystallographic planes~\cite{Li2020}. Intragranular cracks are typically narrow (on the order of tens of nanometers) and may originate at dislocation cores before extending along mechanically weak lattice planes~\cite{Morzy2024}. Under high-voltage operation, this intragranular mode is associated with pronounced surface reactivity, oxygen release, and the formation of a rock-salt surface layer in Ni-rich cathodes~\cite{Xu2021,Jung2017}. Scanning electron microscopy (SEM) and X-ray microscopy can reveal such cracks \emph{ex situ}, after cell disassembly and sample preparation (direct \emph{in situ} imaging at comparable spatial resolution and contrast inside sealed cells is typically impractical). However, translating image contrast into quantitative, class-aware crack maps remains labor-intensive: large SEM images (e.g., $17k \times 7k$~px) must be subdivided into tiles. In our experience, manual labelling can take up to 5~min per tile, amounting to several hours per image and introducing substantial inter-annotator variability~\cite{Lee2023,Hu2023,Chen2025}.

Current imaging methods cannot simultaneously achieve high spatial resolution and track crack evolution over time. High-fidelity techniques such as FIB-SEM or nano-CT resolve fine microstructural details but are destructive, precluding subsequent cycling of the same specimen. Non-destructive modalities such as laboratory micro-CT enable repeated imaging over cycling but typically lack the spatial resolution needed to resolve small cracks. This trade-off motivates approaches that use information from destructive measurements to interpret non-destructive data.

Foundation-model transfer is well suited to this setting. Self-supervised vision-transformer encoders pretrained on natural-image corpora produce transferable mid-level features without domain-specific pretraining; recent encoders~\cite{dinov3_arxiv} preserve dense per-patch correlations and remain effective with a frozen backbone and a lightweight task-specific head~\cite{Karimi2021}. In low-label scientific imaging, where expert annotation is the bottleneck, freezing the encoder concentrates the annotation budget on per-particle morphometric quantification rather than on encoder fine-tuning. The approach therefore fits destructive-microscopy domains in which each annotated tile is costly and the downstream question is morphometric rather than segmentation-overlap based.

The three crack classes correspond to morphological categories associated in the literature with distinct degradation mechanisms in NMC cathodes. Intragranular cracking has been linked to anisotropic lattice strain within primary particles during the high-state-of-charge H2$\rightarrow$H3 phase transition, often aligning with the lamellar (003) structure of the layered oxide~\cite{Yan2017,Xu2021}. Early intergranular cracking propagates along grain boundaries under repeated mechanical stress from volume changes during cycling, while late intergranular cracking represents the developed network state; wider crack openings infiltrated by electrolyte decomposition products and the rock-salt surface layer characteristic of Ni-rich degradation~\cite{Jung2017,Morzy2024}. Resolving these three regimes from a single SEM cross-section therefore captures the chemo-mechanical degradation history at the particle level and supports mechanistic interpretation beyond the binary ``cracked / not cracked'' framing of single-class segmentation.

Deep learning-based segmentation, including variational autoencoder (VAE)-style and U-Net approaches, has accelerated automated analysis of battery microstructures, and recent lithium-plating work suggests that latent representations from autoencoder-style models can encode diagnostically relevant post-mortem image features~\cite{TegetmeyerKleine2026plating}. By contrast, foundation-model encoders pretrained at scale provide richer and more transferable representations for downstream dense prediction~\cite{Oquab2023,dinov3_arxiv,horuseye2026}. However, across the key studies cited here, most contributions either target a single crack morphology or limited volumes (e.g., sub-volume analysis~\cite{Fu2022}), or they primarily advance general segmentation strategies rather than broad, multi-class crack mapping (e.g., multi-phase 3D segmentation~\cite{Mueller2021}, general-purpose segmentation frameworks~\cite{Boyce2022,Majurski2021}, or foundational encoder-decoder architectures~\cite{ronneberger2015unet,xiao2018upernet}). Multi-class crack typing across hundred-megapixel SEM images therefore remains an open problem, and foundation-model transfer pipelines for multi-class segmentation at this resolution in electron microscopy have not been reported. Recent foundation models for segmentation, including the Segment Anything Model (SAM~\cite{kirillov2023sam}) and its successor SAM-2~\cite{ravi2024sam2}, as well as CellPose~\cite{stringer2021cellpose} for biological microscopy, offer zero- or few-shot segmentation via interactive prompting.
However, these models do not natively support class-discriminative multi-label prediction for semantically distinct sub-structures (e.g., early vs.\ late intergranular cracks), and their applicability to narrow, high-aspect-ratio structures in electron microscopy without task-specific fine-tuning remains limited.

Here we introduce a sparse-label morphometric-inference framework that turns frozen foundation-model representations into per-particle crack morphometrics from few expert annotations, making population-scale degradation diagnostics practical for battery-lifetime research. We freeze a self-supervised vision-transformer encoder~\cite{dinov3_arxiv} and train only a lightweight decoder~\cite{deeplabv3plus_arxiv} on the SEM labels. An iterative model-assisted annotation loop grows 79 hand-labelled SEM tiles into a 482-tile expert-corrected reference set across hundred-megapixel cathode cross-sections at three aging states. We resolve intragranular cracking and two stages of intergranular cracking, three categories linked in the literature to distinct degradation mechanisms, and read per-particle width, tortuosity and area fraction from the resulting class maps. Cycled-aged NMC cathodes show an approximately ninefold increase in late intergranular cracking, accompanied by more tortuous, higher-coverage crack networks, whereas calendar-aged samples remain near baseline. These morphometric differences are consistent with degradation associated with repeated cycling rather than elevated-temperature storage alone. Each cross-section yields hundreds of per-particle morphometric measurements, so the scientific findings come from these per-particle distributions rather than from pixel-overlap metrics.

\section{Results}\label{3_Results}

\begin{figure}[htbp]
    \centering
    \includegraphics[width=0.9\linewidth]{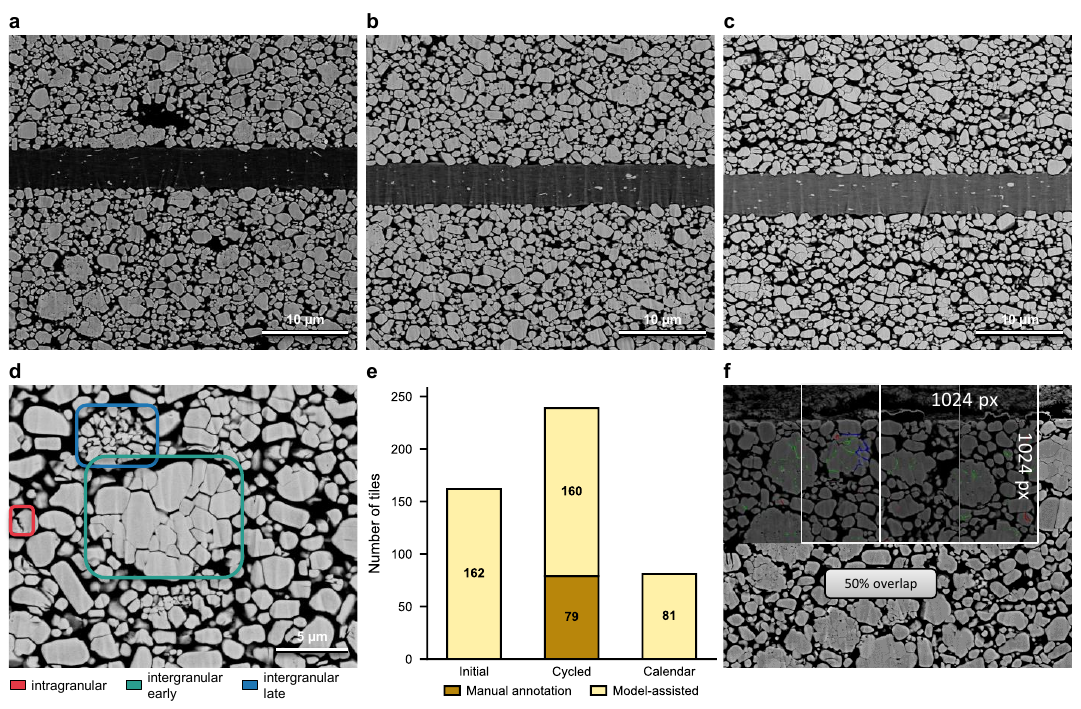}
    \caption{\textbf{Cathode degradation and multi-class crack dataset.}
    (a--c) Representative regions from raw SEM images (ca.\ $17k \times 7k$~px, 22.3~nm/px) of NMC cathode cross-sections: (a) initial, (b) cycled-aged, and (c) calendar-aged. The three samples exhibit visibly different SEM contrast, brightness, and crack density.
    (d) Crack classification scheme: intragranular cracks (within primary grains), early intergranular cracks (typically $<$120~nm at grain boundaries), and late intergranular cracks (typically $>$150~nm and often widened with high-contrast deposits). 
    (e) Dataset composition: 482 expert-corrected tiles distributed as 79 fully manual seed tiles (all drawn from the cycled-aged image) plus 403 model-assisted tiles (160 cycled-aged, 81 calendar-aged, 162 initial), across three samples and crack classes.
    (f) Tiling strategy with 1024$\times$1024~px windows and 50\% overlap for seamless image reconstruction.
    }
    \label{fig:problem_data} 
\end{figure}

We analysed three SEM cross-section images of NMC cathodes representing distinct aging conditions: initial, cycled-aged, and calendar-aged. Each image spans approximately $17k\times7k$~pixels ($382 \times 156$~$\mu$m at 22.3~nm/px). Panels a--c of \cref{fig:problem_data} show representative regions from the three cross-sections and illustrate the condition-dependent differences in SEM contrast, brightness, and crack density. Pixel-level annotations define three crack classes: \emph{early intergranular} (narrow boundary cracks at initial propagation stages, typically $<$120~nm width), \emph{late intergranular} (wider, more developed boundary cracks, typically $>$150~nm width), and \emph{intragranular} (cracks traversing individual grains). The distinction between early and late intergranular cracks is temporal; panel d of \cref{fig:problem_data} summarises the crack taxonomy used throughout the study. In cross-sectional SEM images this progression manifests as crack-width differences, which were assigned by expert visual assessment. Panel f of \cref{fig:problem_data} shows the $1024\times1024$~px tiling strategy with 50\% overlap used for seamless image reconstruction and image-scale inference.
All pixel-level annotations originate from a single expert annotator; systematic biases (e.g., preferential labelling of high-contrast late intergranular cracks over subtle early or intragranular cracks) may affect per-class performance.

\begin{figure}[htbp]
    \centering
    \includegraphics[width=0.9\linewidth]{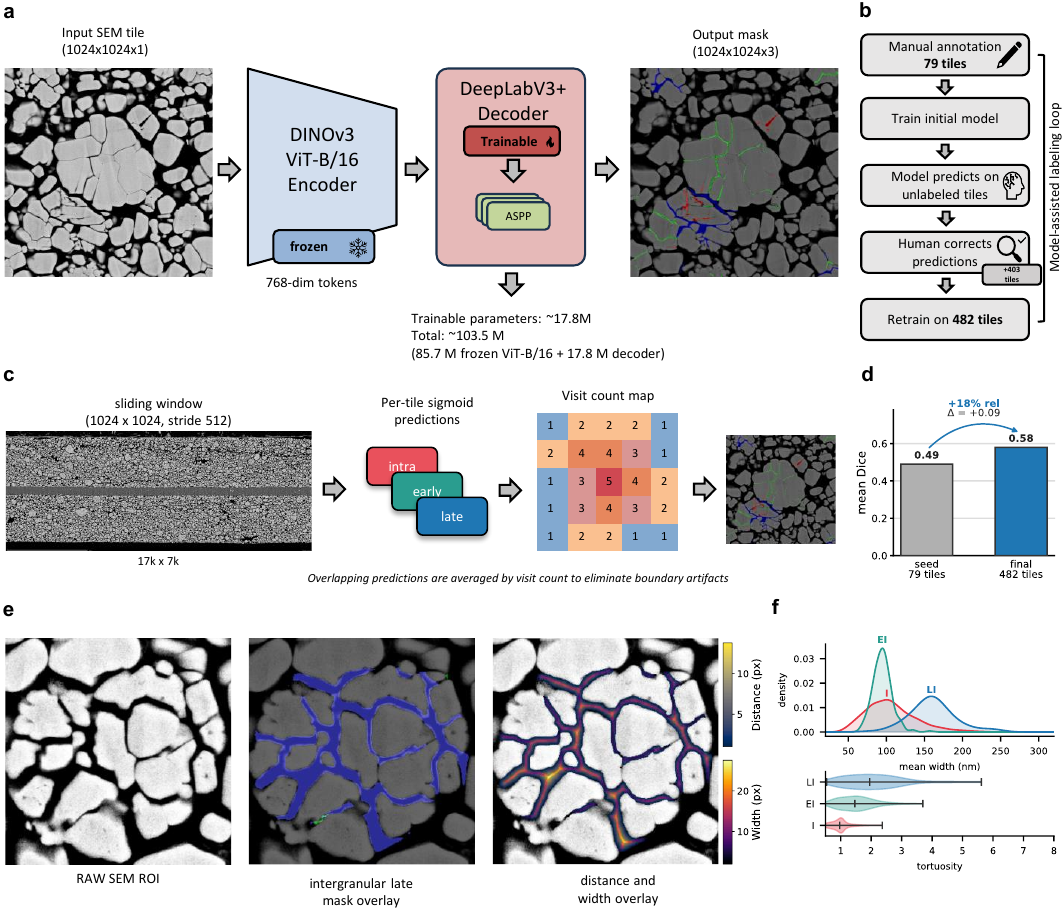}
    \caption{\textbf{Foundation model architecture and model-assisted annotation workflow.}
    (a) Encoder-decoder architecture combining a frozen DINOv3 ViT-B/16 backbone with a trainable DeepLabV3+ decoder for 3-class semantic segmentation. The frozen encoder supplies dense per-patch features while the lightweight decoder adapts to the domain with minimal annotated data and reduced training compute.
    (b) Model-assisted labelling workflow: 79 expert hand-annotated tiles (cycled-aged image) train an initial model. The model then generates provisional pixel-wise crack-class masks for 403 additional tiles (160 cycled-aged, 81 calendar-aged, 162 initial), which are expert-corrected and added back into training to yield the final model.
    (c) Image-scale inference: sliding windows with 50\% overlap produce per-tile 3-class sigmoid predictions (intragranular, early intergranular, late intergranular), stitched by visit-count normalisation to eliminate boundary artefacts.
    (d) Mean-Dice comparison between the seed model (79 manual tiles, 0.49) and the final model (482 expert-corrected tiles, 0.58), an 18\% relative improvement ($\Delta = +0.09$).
    (e) Representative output linking segmentation to geometric quantification: raw SEM ROI, predicted late-intergranular mask overlay, and combined distance- and width-transform overlay (full ROI walkthrough in \sref{fig:si_skeleton}).
    (f) Per-class morphometric summary previewing the full analysis in \cref{fig:morphometric_analysis}: kernel density estimate of mean crack width and tortuosity distributions for intragranular (I), early intergranular (EI), and late intergranular (LI) cracks.}
    \label{fig:method_architecture}
\end{figure}

\begin{figure}[htbp]
    \centering
    \includegraphics[width=0.9\linewidth]{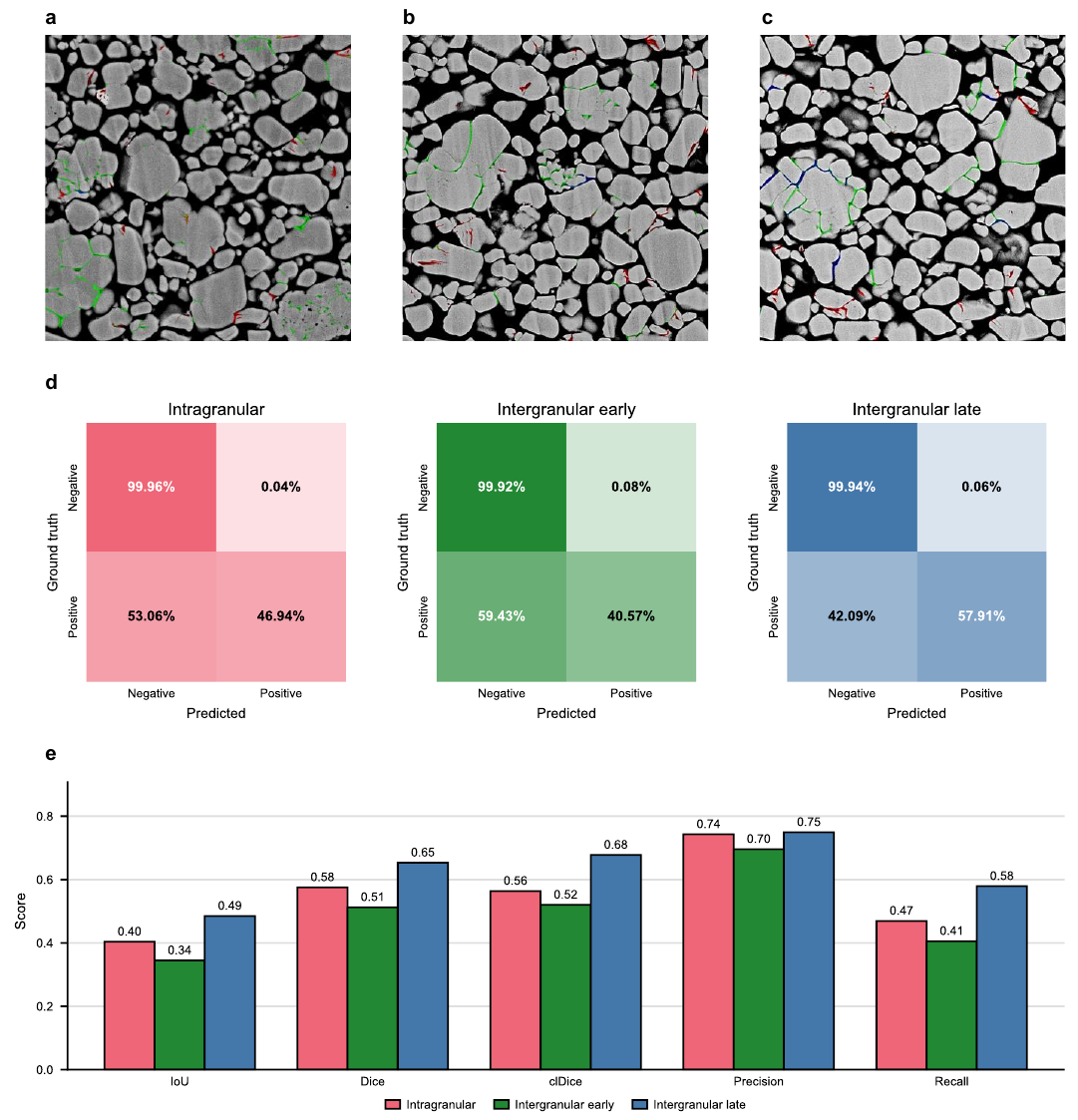}
    \caption{\textbf{Multi-class segmentation performance.}
    (a--c) Representative segmentation results for each sample condition showing input SEM images and model predictions (overlay). Despite the visible contrast and brightness differences among the three SEM images, the model detects the crack classes in all conditions.
    (d) Normalized confusion matrix aggregated across the expert-corrected tile set ($n{=}482$ tiles; threshold $\tau{=}0.1$). 
    (e) Per-class segmentation metrics. The model achieves Dice scores of 0.575 (intragranular), 0.513 (early intergranular), and 0.653 (late intergranular) (mean Dice 0.580; mean clDice 0.587).
    }
    \label{fig:segmentation_results} 
\end{figure}

\subsection*{Multi-class segmentation performance}
We trained the multi-class model using the iterative annotation strategy in \cref{fig:method_architecture}b, with the dataset composition shown in \cref{fig:problem_data}e. An initial model was trained on 79 expert-labelled seed tiles ($1024\times1024$~px each) sampled from the cycled-aged image, achieving a mean Dice score of 0.49. This model was then applied to tiles from all three SEM images to generate provisional pixel-wise crack-class masks, which were expert-corrected and added to the training set. The expanded dataset comprised 482 expert-corrected tiles: 79 fully manual seed tiles (all from the cycled-aged image) and 403 model-assisted tiles distributed across the three SEM cross-sections (160 cycled-aged, 81 calendar-aged, 162 initial), yielding per-sample totals of 239 (cycled-aged), 81 (calendar-aged), and 162 (initial). Retraining on this dataset increased the mean Dice score to 0.58 (\cref{fig:method_architecture}d), an 18\% relative gain from the 79-tile seed model that expanded coverage to all three aging conditions.

We froze a self-supervised vision-transformer encoder (\texttt{facebook/\allowbreak dinov3-\allowbreak vitb16-\allowbreak pretrain-\allowbreak lvd1689m}) and trained only a lightweight decoder (\cref{fig:method_architecture}a). This reduced GPU memory and compute by backpropagating only through the decoder while using transferable self-supervised features~\cite{Karimi2021}. Image-scale inference then followed the sliding-window visit-normalised stitching scheme shown in \cref{fig:method_architecture}c.

On the expert-corrected tile set ($n{=}482$; $\tau{=}0.1$), the final model achieved a mean Dice of 0.580 across the three classes (intragranular: 0.575, early intergranular: 0.513, late intergranular: 0.653; mean precision: 0.729, mean recall: 0.485; \cref{fig:segmentation_results}e). Representative outputs and the aggregated confusion matrix (\cref{fig:segmentation_results}a--d) show that the framework resolves all crack classes across the initial, cycled-aged and calendar-aged SEM cross-sections despite substantial contrast variation between samples. The operating threshold $\tau = 0.1$ was selected from the validation precision--recall curve to maximize class-averaged Dice under the strong foreground--background imbalance of thin crack structures ($<$1\% foreground pixels; \cref{fig:ext_pr_threshold}). Morphometric sensitivity across $\tau \in \{0.05, 0.1, 0.2, 0.3, 0.5\}$ remained qualitatively stable (\cref{fig:ext_tau_sensitivity}), supporting the use of the segmentation outputs for downstream crack morphometrics.

Because many cracks are only 2--5~pixels wide, pixel-exact Dice is conservative: small boundary shifts can substantially reduce overlap without altering the underlying crack morphology~\cite{Joskowicz2019,Reinke2024,Nikolov2021}. Consistent with this, topology-aware centreline Dice (clDice)~\cite{Shit2021clDice} reached a mean of 0.587 (intragranular: 0.564, early intergranular: 0.520, late intergranular: 0.678), while boundary F1 increased from 0.446 at $d = 1$~px tolerance to 0.600 at $d = 2$~px (\sref{fig:si_bf1_distribution}), indicating that residual errors are dominated by small boundary misalignments. The segmentation outputs are therefore used as an intermediate representation for extracting downstream crack morphometrics rather than as the final scientific endpoint.

To assess whether domain-specific supervision is required for crack morphometrics, we evaluated the promptable foundation model SAM2 (Hiera-Large, sam2.1)~\cite{ravi2024sam2} on the same 482-tile reference set using a binary crack-versus-background target (\Cref{tab:extsambaseline}). In automatic mode ($64\times64$ dense point grid, no prompts), SAM2 achieved a binary Dice score of 0.019 [95\% CI 0.018--0.020]. Providing Otsu-derived dark-region centroids as point prompts increased Dice to 0.025 [95\% CI 0.023--0.026], whereas the fine-tuned framework reached a binary Dice score of 0.77 on the same target (paired bootstrap $p < 0.001$, $n = 1000$). Unlike the supervised framework, SAM2 is class-agnostic and therefore cannot directly resolve the intragranular and intergranular crack morphologies required for downstream morphometric analysis (\cref{fig:ext_sam_baseline}; further detail in \sref{fig:si-sam-baseline-zoom}).

\subsection*{Per-particle morphometric quantification}
Per-particle width, tortuosity, and area-fraction distributions extracted from the class maps resolve degradation morphologies that pixel-overlap metrics alone cannot capture.

We automatically identified regions of interest (ROIs) for morphometric analysis from the crack segmentation masks (\sref{fig:si_roi_detection}a--d). For intergranular cracks (early and late), the ROIs approximate particle-scale regions derived from SEM contrast (97 initial, 69 cycled-aged, 49 calendar-aged; each particle-scale ROI yields one early and one late measurement). Intragranular ROIs correspond to connected crack components within individual particles (446 initial, 610 cycled-aged, 623 calendar-aged). Within each ROI, crack centrelines were extracted by skeletonization and local widths were computed from the Euclidean distance transform (\sref{fig:si_skeleton}c--f; previewed in \cref{fig:method_architecture}e). Each crack class was characterised by three skeleton-based descriptors (see Methods for definitions and \Cref{tab:extmorphometrics} for summary statistics): area fraction (percentage of ROI area occupied by crack pixels), mean width (average maximal-inscribed-disk diameter along the centreline), and geometric tortuosity (centreline arc length divided by the projected span along the principal crack axis).

Cycling strongly increases late-stage intergranular cracking relative to both the initial and calendar-aged states, consistent with progressive grain-boundary degradation during repeated electrochemical cycling. The cycled-aged cross-section shows a late intergranular area fraction of $4.58\%$, compared with $0.52\%$ in the initial sample and $0.49\%$ in the calendar-aged sample, corresponding to an approximately ninefold increase relative to the initial state. In contrast, the calendar-aged sample remains near baseline despite elevated-temperature storage. Early intergranular coverage remains comparatively stable across conditions ($1.30$--$2.34\%$), with modest width differences ($\sim$91--99\,nm), consistent with a broadly similar early-stage boundary-opening morphology.

\begin{figure}[htbp]
  \centering
  \includegraphics[width=0.9\linewidth]{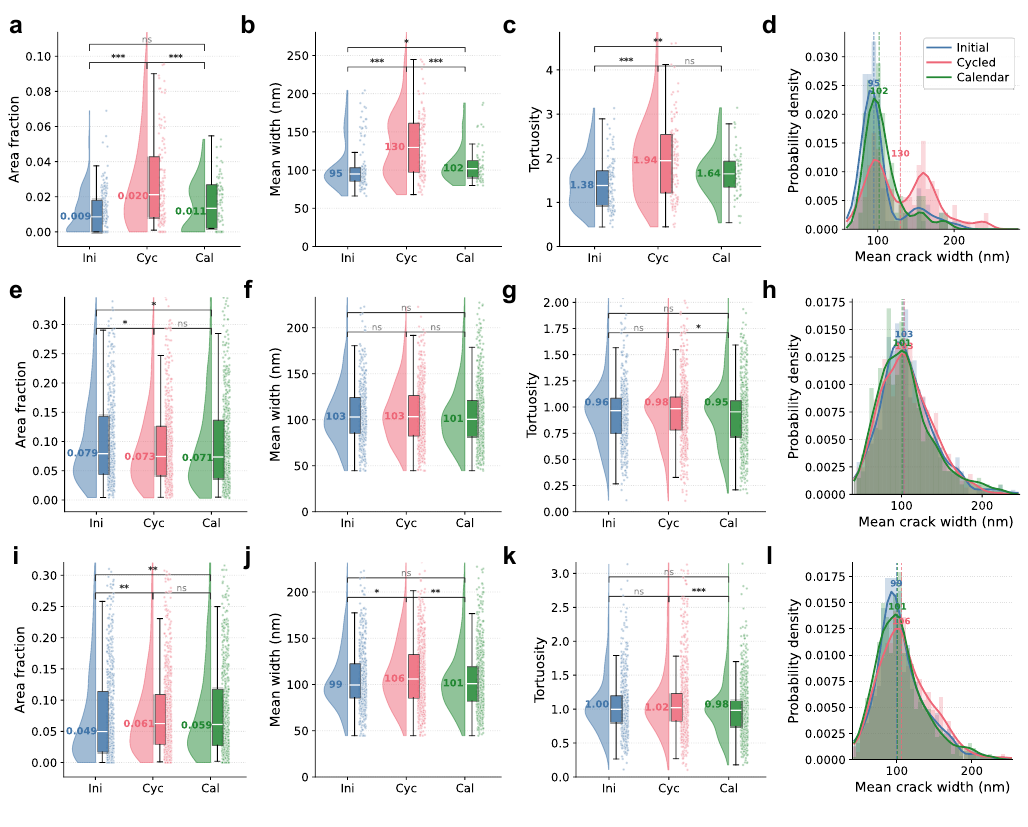}
  \caption{\textbf{Automated morphometric quantification of crack characteristics.}
    Three rows show intergranular cracks (a--d, combining early and late classes), intragranular cracks (e--h), and all cracks combined (i--l).
    Each row contains four panels:
    violin plots of area fraction (a,\,e,\,i),
    mean crack width (b,\,f,\,j),
    tortuosity (c,\,g,\,k),
    and width probability distributions (d,\,h,\,l).
    Brackets indicate statistically significant pairwise differences (two-sided Mann--Whitney $U$ test);
    significance levels: ***~$q<0.001$, **~$0.001\le q<0.01$, *~$0.01\le q<0.05$, ns~$q\ge0.05$, where $q$ is the Benjamini--Hochberg false-discovery-rate-adjusted $p$-value across the family of 27 pairwise tests (3 subsets $\times$ 3 metrics $\times$ 3 condition pairs); see Methods.
    Cycled-aged samples show significantly higher late intergranular crack area fraction and tortuosity than initial and calendar-aged samples; mean crack width is only slightly larger and not statistically significant (panel~j).}
  \label{fig:morphometric_analysis}
\end{figure}

Late intergranular cracks in the cycled-aged sample are markedly more tortuous than those in the initial and calendar-aged samples, indicating the development of more interconnected grain-boundary crack networks during electrochemical cycling. Mean late-crack width increases only modestly from $159 \pm 22$~nm in the initial sample to $167 \pm 39$~nm in the cycled-aged sample (not statistically significant; \cref{fig:statistical_analysis}j), whereas tortuosity increases significantly from $1.55 \pm 0.63$ to $2.40 \pm 1.00$. ROI-level distributions of area fraction, mean width, tortuosity, and per-pixel width probability density are summarised in \cref{fig:morphometric_analysis}a--d (previewed in \cref{fig:method_architecture}f). Cycled-aged late cracks occupy a broader range of area fractions, widths, and tortuosities than the initial and calendar-aged samples, with pairwise two-sided Mann--Whitney $U$ tests indicating statistically significant distributional shifts and Cliff's $\delta$ values supporting substantial effect sizes. Intragranular morphometrics (\cref{fig:morphometric_analysis}e--h) remain comparatively similar across conditions, whereas combined-class distributions (\cref{fig:morphometric_analysis}i--l) retain the cycled-aged shift toward larger widths and tortuosities, including a broader width tail extending beyond 200\,nm for late intergranular cracks. These condition-level trends are preserved across threshold reruns (\cref{fig:ext_tau_sensitivity}).

To summarise condition-dependent morphometric differences, we compared ROI-level distributions using two-sided Mann--Whitney $U$ tests with Cliff's $\delta$ effect sizes (see Methods). ECDFs, orientation roses, and area-fraction-versus-width scatter plots for intergranular, intragranular, and combined populations are shown in \cref{fig:statistical_analysis}a--i, with the intergranular Mann--Whitney $U$ and Cliff's $\delta$ summary in \cref{fig:statistical_analysis}j. For late intergranular area fraction, cycled-aged ROIs exceeded initial ROIs ($U=586$, $q<0.001$, $\delta=-0.82$), indicating a strong shift toward higher crack coverage under cycling, whereas the calendar-aged distribution remained similar to the initial state ($U=2{,}419$, $q=0.85$, $\delta=+0.02$). Because each condition is represented by a single cross-section, these ROI-level statistics describe within-cross-section morphometric heterogeneity rather than cell- or lot-level population inference (see Discussion). Additional validation is provided by \cref{fig:ext_cross_sample} and \sref{fig:si_metric_vs_density,fig:si_bf1_distribution}.



\definecolor{headgray}{RGB}{230,230,230}
\definecolor{groupA}{RGB}{245,248,252}   
\definecolor{groupB}{RGB}{248,220,220}   

\newcommand{\FigInnerWidth}{0.95\linewidth} 

\begin{figure}[htbp]
  \centering

  \includegraphics[width=0.9\linewidth]{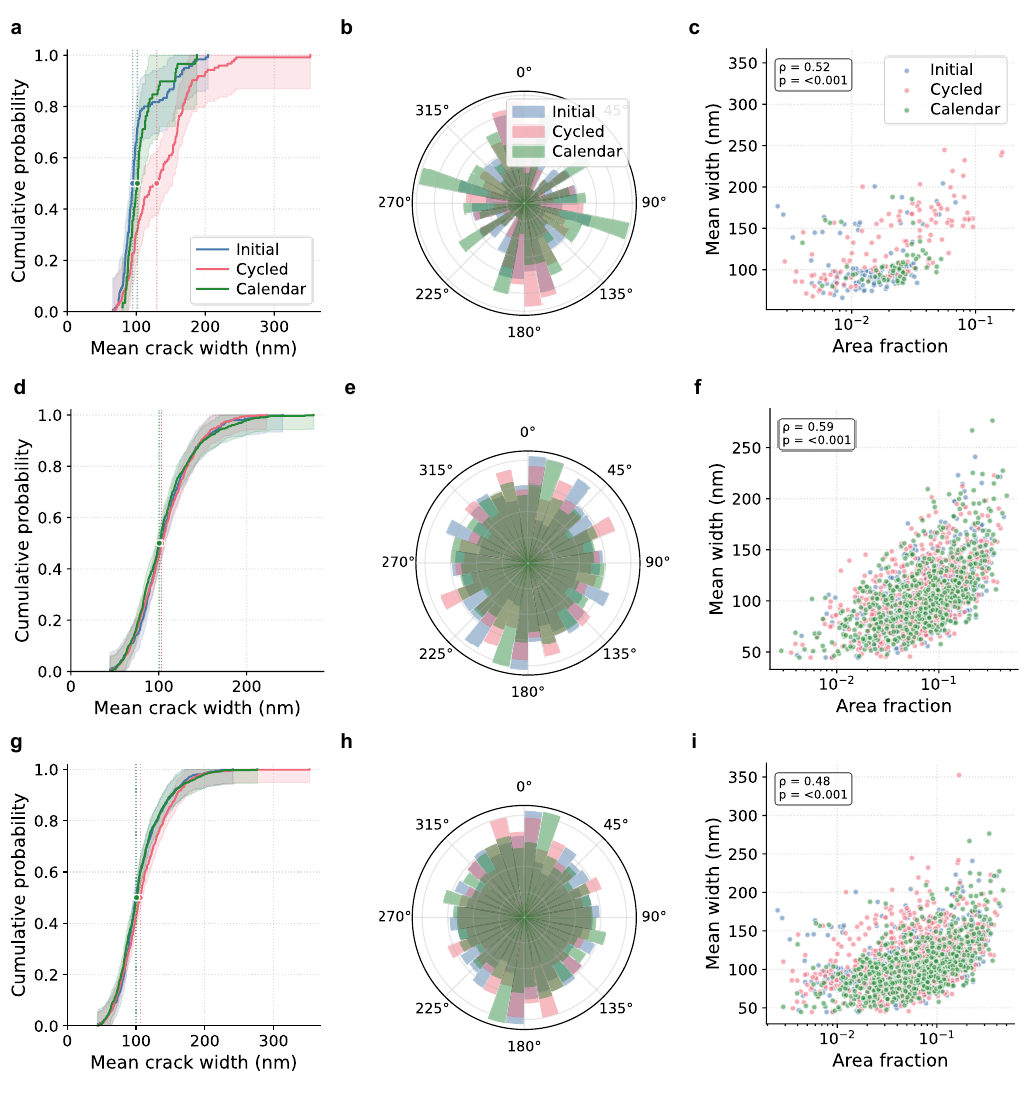}

  \vspace{0.8em}

\begin{minipage}{\FigInnerWidth}
  {\fontfamily{phv}\bfseries j}\par\vspace{0.2em}
  \centering
  \renewcommand{\arraystretch}{0.15}
  \setlength{\tabcolsep}{4pt}
    {\small
    \resizebox{\FigInnerWidth}{!}{%
    \begin{tabular}{@{}l l r c r c@{}}
      \toprule
      \rowcolor{headgray}
      \textbf{Comparison} & \textbf{Metric} &
      \cellcolor{groupA}\textbf{U-statistic} &
      \cellcolor{groupA}\textbf{$q$-value} &
      \cellcolor{groupB}\textbf{Cliff's $\delta$} &
      \cellcolor{groupB}\textbf{Effect size} \\
      \midrule

      Ini vs Cyc & Area fraction & \cellcolor{groupA}586 & \cellcolor{groupA}$<0.001$ & \cellcolor{groupB}$-0.82$ & \cellcolor{groupB}Large \\
      Ini vs Cal & Area fraction & \cellcolor{groupA}2{,}419 & \cellcolor{groupA}0.85 & \cellcolor{groupB}$+0.02$ & \cellcolor{groupB}Negligible \\
      Cyc vs Cal & Area fraction & \cellcolor{groupA}3{,}098 & \cellcolor{groupA}$<0.001$ & \cellcolor{groupB}$+0.83$ & \cellcolor{groupB}Large \\
      \addlinespace

      Ini vs Cyc & Width (nm) & \cellcolor{groupA}709 & \cellcolor{groupA}0.60 & \cellcolor{groupB}$-0.09$ & \cellcolor{groupB}Negligible \\
      Ini vs Cal & Width (nm) & \cellcolor{groupA}160 & \cellcolor{groupA}0.44 & \cellcolor{groupB}$+0.21$ & \cellcolor{groupB}Small \\
      Cyc vs Cal & Width (nm) & \cellcolor{groupA}482 & \cellcolor{groupA}0.15 & \cellcolor{groupB}$+0.35$ & \cellcolor{groupB}Medium \\
      \addlinespace

      Ini vs Cyc & Tortuosity & \cellcolor{groupA}352 & \cellcolor{groupA}$<0.001$ & \cellcolor{groupB}$-0.55$ & \cellcolor{groupB}Large \\
      Ini vs Cal & Tortuosity & \cellcolor{groupA}143 & \cellcolor{groupA}0.77 & \cellcolor{groupB}$+0.08$ & \cellcolor{groupB}Negligible \\
      Cyc vs Cal & Tortuosity & \cellcolor{groupA}565 & \cellcolor{groupA}0.010 & \cellcolor{groupB}$+0.58$ & \cellcolor{groupB}Large \\

      \bottomrule
    \end{tabular}%
    }%
    }
  \end{minipage}

  \caption{\textbf{Statistical analysis of degradation patterns.}
    Rows: intergranular (a--c), intragranular (d--f), and all cracks combined (g--i). Columns: ECDFs of mean crack width with 95\% Dvoretzky--Kiefer--Wolfowitz bands and median markers (a,\,d,\,g); orientation rose diagrams (b,\,e,\,h); area fraction vs.\ mean width (c,\,f,\,i).
    The cycled-aged width distributions (red) show only a slight, non-significant shift toward larger widths relative to initial (blue) and calendar-aged (green); the statistically significant cycled-aged increases are in area fraction and tortuosity (panel~j).
    (j) Mann--Whitney $U$ tests on late intergranular cracks with Cliff's $\delta$; $q$-values are Benjamini--Hochberg-adjusted within the 9-test intergranular-late sub-family (see Methods and Supplementary Methods~\S S7).
  }
  \label{fig:statistical_analysis}
\end{figure}

\section{Discussion}\label{4_Discussion}

\begin{figure}[!t]
  \centering
  \includegraphics[width=0.9\linewidth]{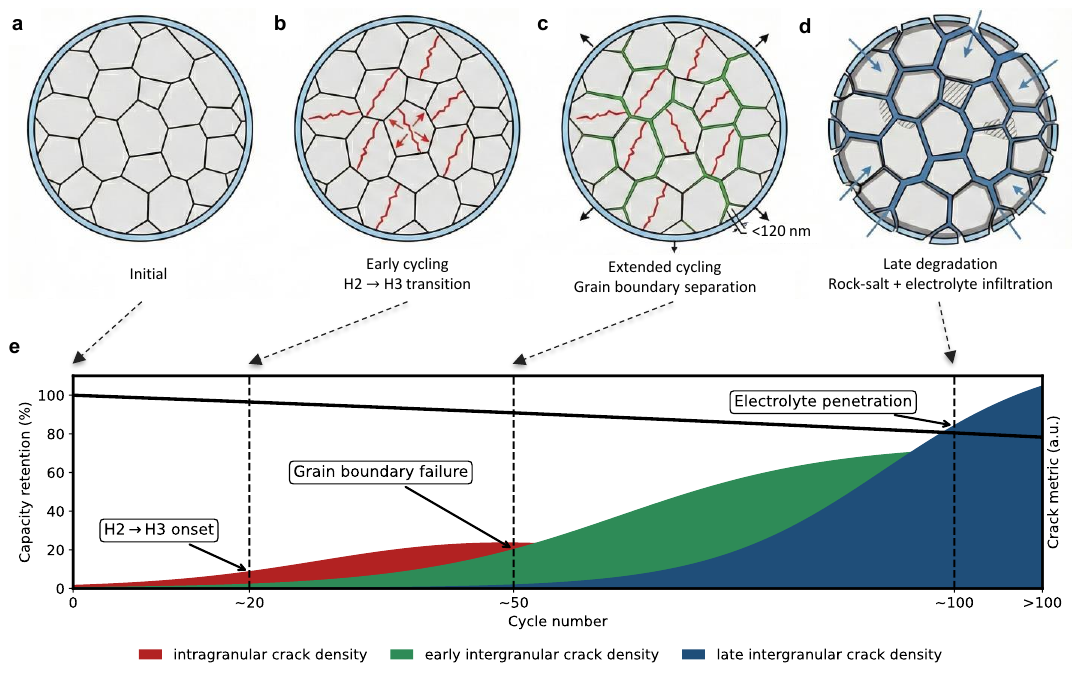}
  \caption{\textbf{Proposed mechanism of crack evolution in NMC cathodes.}
    (a--d) Schematic illustration of characteristic crack types, shown as a generalized progression; in practice, these mechanisms may overlap temporally and coexist depending on local microstructural conditions:
    (a) Initial polycrystalline NMC particle with coherent grain boundaries.
    (b) Intragranular cracks form within primary particles due to anisotropic lattice strain during H2$\rightarrow$H3 phase transition.
    (c) Intergranular cracks nucleate at grain boundaries under repeated mechanical stress from volume changes. Intragranular cracks from~(b) persist but are omitted for visual clarity.
    (d) Late-stage degradation with widened cracks ($>$150~nm), rock-salt phase formation, and electrolyte infiltration pathways. Intragranular cracks remain present but are not shown to highlight intergranular crack progression.
    (e) Generalized trend of crack morphology metrics correlated with electrochemical capacity retention, suggesting crack width as a quantitative degradation marker. The depicted sequence represents a predominant tendency rather than a strict temporal order.}
  \label{fig:mechanistic_interpretation}
\end{figure}

Cycling produces a strong enrichment of late intergranular cracking in NMC that is not observed in the calendar-aged sample under elevated-temperature storage, consistent with progressive grain-boundary degradation during repeated electrochemical lithiation--delithiation~\cite{Pender2020,Edge2021}. The accompanying increase in crack area fraction and geometric tortuosity (\Cref{tab:extmorphometrics}) indicates that cycling is associated with both greater grain-boundary crack coverage and the development of more interconnected crack networks relative to the initial and calendar-aged states.

The morphologies observed in the cycled-aged samples are consistent with the anisotropic lattice evolution of Ni-rich NMC during electrochemical cycling, in which the in-plane lattice contracts while the c-axis undergoes expansion followed by collapse near the H2$\rightarrow$H3 transition at high state of charge~\cite{Kondrakov2017,Ryu2018,debiasi2019}. Such anisotropic deformation has been associated with mechanically incompatible strain between differently oriented primary grains and the development of intergranular cracking~\cite{Yan2018grain,heenan2020,chemomech2024}. The more tortuous, higher-coverage intergranular crack networks observed here are also consistent with progressive grain-boundary degradation and surface reconstruction reported in Ni-rich cathodes, including rock-salt formation and lattice-oxygen-related surface reactivity~\cite{Yan2017,Jung2017,Xu2021,Morzy2024}. The accompanying increase in geometric tortuosity ($2.40 \pm 1.00$ vs.\ $1.55 \pm 0.63$) and mean crack length ($14.2 \pm 10.5$~$\mu$m vs.\ $1.5 \pm 3.4$~$\mu$m) further supports the development of increasingly interconnected crack networks under cycling~\cite{Li2020,Brandt2020,Britala2023}. In contrast, the calendar-aged sample shows comparatively limited crack evolution despite elevated-temperature storage, consistent with the absence of repeated electrochemical strain~\cite{Du2021,Mao2017,Jung2017gas}. \Cref{fig:mechanistic_interpretation} summarises the proposed progression of crack-network development and the associated morphometric trends. The skeleton-based tortuosity reported here describes individual crack-path geometry rather than effective pore-network transport tortuosity.

The methodological contribution of this work is to extract per-particle crack morphometrics from sparse and imperfect SEM annotations, concentrating the expert-label budget on the scientific question rather than on representation learning. Per-pixel class probabilities serve as an intermediate representation from which distributions of crack width, tortuosity and area fraction are derived. By freezing a self-supervised encoder~\cite{dinov3_arxiv} and training only a lightweight decoder~\cite{deeplabv3plus_arxiv}, the framework leverages transferable visual representations despite the limited number of expert-corrected SEM tiles. This extends transfer-learning approaches in low-data scientific imaging~\cite{Karimi2021,umedpt2024} to multi-class morphometric analysis at the hundred-megapixel scale, beyond earlier battery-microstructure studies focused on single crack classes or smaller imaging volumes~\cite{Mueller2021,Fu2022,ronneberger2015unet}. The morphometric formulation is also relatively tolerant to local annotation uncertainty because width and tortuosity measurements average over crack geometry rather than relying on exact pixel overlap. Consistent morphometric trends across thresholds and descriptors further support the robustness of the extracted degradation signatures beyond pixel-level segmentation metrics.

Each destructive cross-section yields hundreds of per-particle morphometric measurements, enabling population-level characterization of crack morphology from sparse SEM annotation. More broadly, sparse-label morphometric inference may be applicable to other destructive-microscopy domains in which annotation cost limits throughput, including ceramic microstructure, geological thin-sections, polymeric composites, and biological tissues. Self-supervised encoders~\cite{dinov3_arxiv} and promptable segmenters~\cite{kirillov2023sam,ravi2024sam2} have begun to demonstrate broad transfer in scientific imaging; the contribution here is to make that transfer quantitative in a morphometric-analysis setting rather than as a segmentation benchmark alone. Standardised morphology-based diagnostics built on this framework would support lithium-ion manufacturing quality assurance, aging diagnostics, and quantification of single-crystal NMC designs that suppress intergranular cracking~\cite{Li2017single,Qian2020,gradporous2024}. Extensions to 3D imaging (XRM, nano-CT), multimodal fusion (EDS, EBSD), active-learning annotation, and coupling with electrochemical diagnostics for physics-based aging models are natural next steps.

Several limitations of the present case study should be noted. First, the focus of this work is the sparse-label morphometric-inference framework rather than optimisation of a specific segmentation architecture; the frozen vision-transformer encoder with a lightweight decoder represents one implementation of this approach. Systematic benchmarking across alternative self-supervised encoders, decoder architectures, and training seeds is reserved for future work. Second, only three SEM cross-sections (one per aging condition) were available because high-resolution destructive imaging is costly, so the reported ROI-level statistics describe within-cross-section morphometric heterogeneity rather than cell- or lot-level population inference. Third, two-dimensional sectioning cannot fully resolve out-of-plane crack morphology or pore--crack ambiguity, motivating future volumetric validation using FIB-SEM or X-ray microscopy. Finally, the framework was trained on NMC images acquired on a single instrument; extension to other cathode chemistries, imaging conditions or laboratories will likely require domain adaptation and larger multi-site reference datasets.

\section{Conclusions}
We introduced a sparse-label morphometric-inference framework that turns frozen self-supervised vision-transformer features into per-particle crack morphometrics from few expert annotations, concentrating the annotation budget on quantification rather than representation learning. An iterative model-assisted annotation loop grew 79 hand-labelled tiles into a 482-tile expert-corrected reference set across three hundred-megapixel NMC cathode cross-sections, resolving intragranular as well as early- and late-stage intergranular cracking. The cycled-aged cathode exhibited an approximately ninefold increase in late intergranular cracking and more tortuous, higher-coverage crack networks, whereas the calendar-aged sample remained near baseline, consistent with degradation from repeated electrochemical cycling rather than elevated-temperature storage alone. These condition-level trends were preserved across decision thresholds and topology-aware metrics, while a zero-shot SAM2 baseline failed on the thin crack structures, indicating that domain-specific supervision is needed for this task. By converting a single destructive cross-section into population-level crack statistics, the framework supports aging diagnostics, manufacturing quality assurance and second-life assessment, and extends naturally to other destructive-microscopy domains in which annotation cost limits throughput.

\section{Methods}\label{Methods}

\subsection*{Multi-class segmentation of crack types}
Pixel-wise crack typing is cast as a three-class semantic segmentation problem with labels \emph{intragranular} (reflecting defect-induced crack initiation within primary particles~\cite{Lin2020}), \emph{early intergranular}, and \emph{late intergranular}.

The frozen self-supervised DINOv3 encoder~\cite{dinov3_arxiv} provides dense texture- and morphology-aware features. The checkpoint \texttt{facebook/\allowbreak dinov3-\allowbreak vitb16-\allowbreak pretrain-\allowbreak lvd1689m} is a ViT-B/16 backbone (85.7~M frozen parameters, 768-dimensional patch embeddings). Each greyscale $1024\times1024$~px SEM tile is replicated to three channels and tokenised on a $64\times64$ grid of $16\times16$ patches, yielding a $64\times64\times768$ feature map that is passed to a DeepLabV3+ decoder~\cite{deeplabv3plus_arxiv} (17.8~M trainable parameters). For an input image $I\in[0,1]^{H\times W}$, the network predicts per-pixel logits $\ell\in\mathbb{R}^{C\times H\times W}$ with $C{=}3$; per-class probabilities use independent sigmoids,

\begin{equation}
p_{c}(x) \;=\; \sigma\!\big(\ell_c(x)\big)
\;=\; \frac{1}{1+\exp\!\big(-\ell_c(x)\big)},
\quad c\in\{1,\ldots,C\},\; x\in\Omega,
\label{eq:sigmoid}
\end{equation}
and binary masks follow by thresholding, $\hat{y}_c(x)=\mathbf{1}[\,p_c(x)\ge\tau\,]$. The operating threshold $\tau{=}0.1$ was selected on the validation precision--recall curve (architecture in \cref{fig:method_architecture}a; sensitivity in \cref{fig:ext_tau_sensitivity}). Training minimised a weighted sum of binary cross-entropy, mean-squared-error, and Dice losses (Supplementary Methods~\S S1).

Tiles from each cross-section were randomly partitioned 80\%/20\% per image, yielding 96 held-out validation tiles from a 482-tile pool: 79 manual seed tiles (cycled-aged) and 403 model-assisted, expert-corrected tiles distributed as 160 (cycled-aged), 81 (calendar-aged), and 162 (initial). Because tiles share 50\% spatial overlap by construction, pixel-level segmentation metrics are reported on the full 482-tile expert-corrected reference as operating-point calibration rather than out-of-distribution generalisation (Supplementary Methods~\S S5). The ROI-level morphometric distributions reported in the Results are computed on the full cross-sections and are independent of the tile-level split.

\subsection*{Baseline evaluation using SAM2}
The Segment Anything Model 2 (\texttt{facebook/sam2.1-hiera-large}) was evaluated in two configurations: automatic mask generation on a dense point grid, and prompted segmentation using Otsu-thresholded dark regions as point prompts. Because SAM2 is class-agnostic, the comparison was performed on a binary crack-versus-background target obtained by collapsing the three reference classes; on this binary metric the DINOv3 framework reaches Dice $=0.77$ against the SAM2 values reported in \cref{tab:extsambaseline} and visualised in \cref{fig:ext_sam_baseline}. Prompt-grid spacing, intensity and geometric post-filters, and multi-mask selection are described in Supplementary Methods~\S S6.

\subsection*{Inference and overlay mask generation}
Inference on full-resolution images ($\sim 17{,}000 \times 7{,}000$~px) used an overlapping sliding-window strategy, averaging per-class predictions across overlapping tiles (Supplementary Methods~\S S2). Final per-class binary masks were thresholded and rendered as alpha-blended overlays at native resolution for side-by-side comparison across the \emph{initial}, \emph{calendar-aged}, and \emph{cycled-aged} probes.

\subsection*{Segmentation mask analysis pipeline}
The morphometric pipeline transforms each per-pixel likelihood map into a cleaned binary crack mask, extracts particle-level regions of interest (ROIs) corresponding to individual cathode secondary particles (or small clusters), and reduces each ROI to skeleton-based width and geometry distributions; all measurements are computed in pixel units and converted to physical units via SEM calibration. The crack likelihood map $m(x)\in[0,1]$ is thresholded at $\tau$ to give a binary field $B(x)=\mathbf{1}\{m(x)\ge\tau\}$, with small connected components filtered by area and skeleton length, and ROIs are obtained from the SEM backscatter contrast (panels a--e of \sref{fig:si_roi_detection}; full procedure in Supplementary Methods~\S S2).

Within each ROI, the binary crack mask is reduced to one-pixel centrelines by skeletonisation, $S\subset\Omega$. The Euclidean distance transform on the crack boundary $\partial B$ (panels c, d of \sref{fig:si_skeleton}),
\begin{equation}
d(x)\;=\;\min_{y\in \partial B}\,\|x-y\|_2,
\label{eq:edt}
\end{equation}
defines a local crack width at each skeleton point $x_s\in S$,
\begin{equation}
w(x_s)\;=\;2\,d(x_s),
\label{eq:width}
\end{equation}
the diameter of the maximal inscribed disk centred at $x_s$. Non-finite or non-positive samples are excluded from summary statistics but retained as zeros in visualisation maps.

\subsection*{Geometric descriptors and width distributions}
Panel c of \sref{fig:si_skeleton} illustrates orientation, topology, and tortuosity for an example ROI.
Let $\{x_i\}_{i=1}^{n}$ be skeleton coordinates in image axes. We summarise:
\begin{itemize}
  \item \textbf{Crack length:} $L=\sum_{x\in S} 1$ (in pixels).
  \item \textbf{Area fraction:} $| \{x: B(x)=1\} | / |\Omega_{\mathrm{ROI}}|$.
  \item \textbf{Topology:} classify pixels by $3\times3$ neighbour degree (endpoints: degree 1; branch points: degree $\ge3$; isolated: degree 0).
  \item \textbf{Orientation and anisotropy:} with centered coordinates $X=[x_i-\bar{x}]$, the covariance $\Sigma=\frac{1}{n-1}X^\top X$ has eigenpairs $(\lambda_1,\mathbf{v}_1)$, $(\lambda_2,\mathbf{v}_2)$, $\lambda_1\ge\lambda_2\ge0$. Orientation is $\angle\mathbf{v}_1\in[0^\circ,180^\circ)$ and anisotropy is
  \begin{equation}
  \mathcal{A} \;=\; \frac{\lambda_1}{\lambda_2+\varepsilon}.
  \label{eq:anisotropy}
  \end{equation}
  \item \textbf{Geometric tortuosity:} ratio of centreline length to projected span along $\mathbf{v}_1$; a morphological descriptor, distinct from the effective transport tortuosity of the pore network:
  \begin{equation}
  \mathcal{T} \;=\; \frac{L}{\max_i \langle x_i-\bar{x}, \mathbf{v}_1\rangle - \min_i \langle x_i-\bar{x}, \mathbf{v}_1\rangle + \varepsilon }.
  \label{eq:tortuosity}
  \end{equation}
  Because $L$ is a raw skeleton-pixel count rather than a $\sqrt{2}$-weighted arc length, near-linear cracks can yield $\mathcal{T}$ marginally below unity; such values reflect near-straight morphology rather than a measurement error.
  \item \textbf{Transverse roughness:} RMS deviation orthogonal to $\mathbf{v}_1$,
  \begin{equation}
  \rho \;=\; \sqrt{\frac{1}{n}\sum_{i=1}^{n} \langle x_i-\bar{x}, \mathbf{v}_2\rangle^{2}}.
  \label{eq:roughness}
  \end{equation}
\end{itemize}

Let $\{w_i\}_{i=1}^{N}$ be valid widths sampled on $S$. We report descriptive statistics (min, max, mean, median, std) and visualize: (i) a histogram of $\{w_i\}$ and (ii) the empirical cumulative distribution function (ECDF), defined as
\begin{equation}
\widehat{F}(t)\;=\;\frac{1}{N}\sum_{i=1}^{N}\mathbf{1}\{w_i \le t\},
\label{eq:ecdf}
\end{equation}
i.e.\ the fraction of the centreline length with width $\le t$~\cite{Wasserman2004}. Quartiles ($P_{25},P_{50},P_{75}$) are marked for cross-ROI comparison (panel f of \sref{fig:si_skeleton}; aggregated ECDFs in panels a, d, and g of \cref{fig:statistical_analysis}).

\subsection*{Statistical analysis}
Per-ROI morphometric distributions (area fraction, mean width, geometric tortuosity) were compared between aging conditions using two-sided Mann--Whitney $U$ tests with Cliff's $\delta$ as a non-parametric effect-size estimator and Benjamini--Hochberg correction for family-wise multiple comparisons. Test-family structure, raw versus adjusted statistics, and the released TSV summary are detailed in Supplementary Methods~\S S7.

\subsection*{Reporting and assumptions}
Quantitative measurements derive from thresholded crack masks; reported quantities are in pixels and converted to physical units via SEM calibration. Width statistics depend on $\tau$, with sensitivity reported in \cref{fig:ext_tau_sensitivity}. Cell construction, aging protocols, post-mortem sample preparation, and SEM acquisition parameters are summarised in the Supplementary Information (Sample preparation and SEM imaging).

\section*{CRediT authorship contribution statement}
\textbf{Thorsten Tegetmeyer-Kleine:} Conceptualization, Methodology, Software, Validation, Formal analysis, Data curation, Visualization, Writing -- original draft. \textbf{Thomas Schmitt:} Conceptualization, Writing -- review \& editing. \textbf{Phillip Aquino:} Conceptualization, Writing -- review \& editing. \textbf{Christiane Rahe:} Investigation. \textbf{Dirk Uwe Sauer:} Resources, Writing -- review \& editing. \textbf{Weihan Li:} Supervision, Funding acquisition, Writing -- review \& editing. All authors reviewed and approved the final manuscript.

\section*{Declaration of competing interest}
The authors declare that they have no known competing financial interests or personal relationships that could have appeared to influence the work reported in this paper.

\section*{Acknowledgements}
This work was supported by Honda Research Institute Europe.

\section*{Data availability}
The datasets generated and analysed during this study, including the expert-corrected SEM tile image--mask pairs, SEM cross-sections, released model weights, validation artefacts, and morphometric analysis tables, are openly available on Figshare at \url{https://doi.org/10.6084/m9.figshare.32975921}.

\section*{Code availability}
The code used for annotation support, inference, validation, morphometric analysis, and figure generation in this study, together with the trained model weights, is openly available at \url{https://git.rwth-aachen.de/thorsten.tegetmeyer-kleine/hri-nmc-crack-segmentation}.

\clearpage
\appendix
\section{Extended data}
\label{app:ed}

\setcounter{figure}{0}
\setcounter{table}{0}
\crefalias{table}{exttable}
\renewcommand{\theHfigure}{ED.\arabic{figure}}
\renewcommand{\theHtable}{ED.\arabic{table}}

\begin{figure}[htbp]
  \centering
  \includegraphics[width=0.9\linewidth]{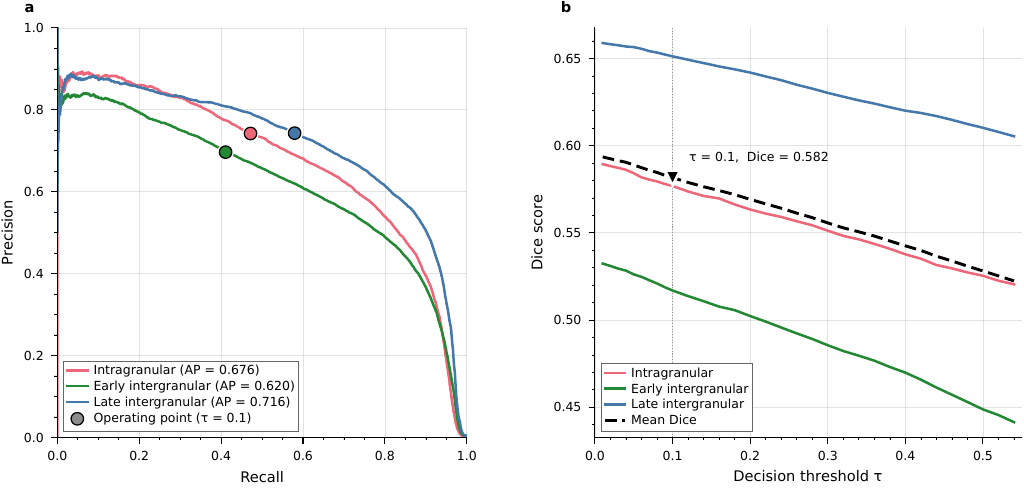}
  \caption{\textbf{$|$ Precision--recall curves and threshold selection.} (a)~Per-class PR curves with average precision (AP) values; filled circles mark the operating point at $\tau{=}0.1$. (b)~Per-class and mean Dice score vs.\ decision threshold~$\tau$; the selected threshold is annotated. The low optimal $\tau$ reflects extreme foreground--background imbalance ($<$1\% crack pixels).}
  \label[extfigure]{fig:ext_pr_threshold}
\end{figure}

\begin{figure}[htbp]
  \centering
  \includegraphics[width=0.9\linewidth]{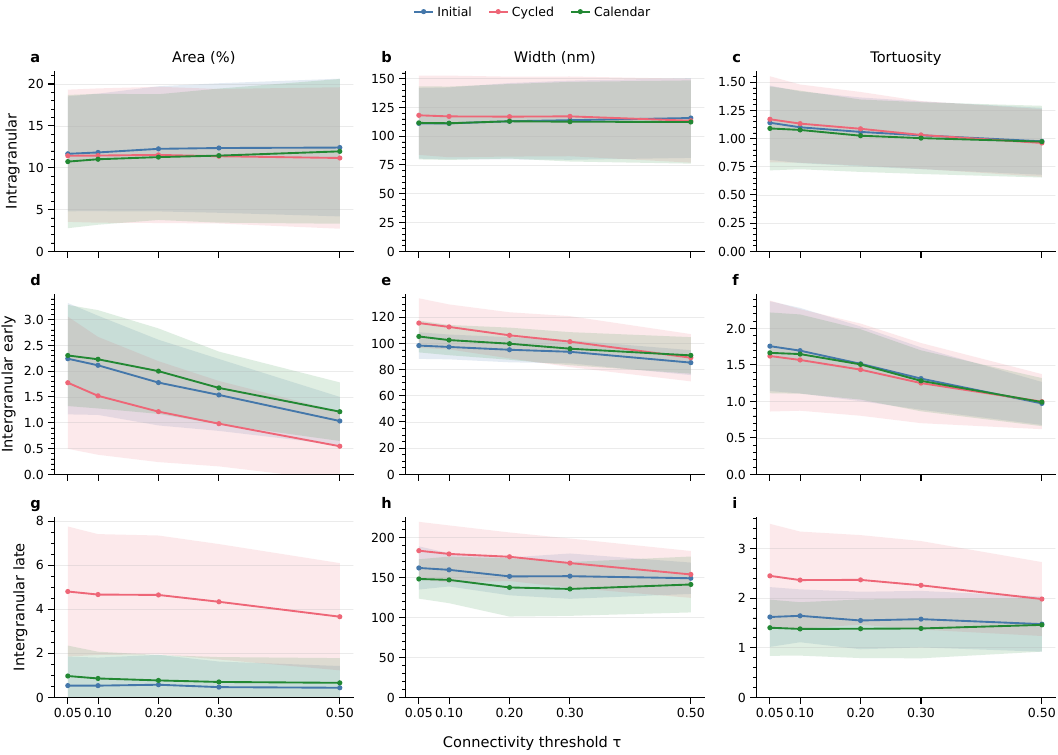}
  \caption{\textbf{$|$ Morphometric threshold sensitivity.} Area fraction, mean crack width, and tortuosity (mean $\pm$ std across ROIs) as a function of the decision threshold $\tau\in\{0.05,0.1,0.2,0.3,0.5\}$ for each crack class and sample condition. Area fraction decreases monotonically with increasing $\tau$ as stricter thresholds suppress low-confidence crack pixels; mean width and tortuosity are more stable because skeleton-based statistics average over retained crack pixels. The relative ordering of conditions is preserved across all $\tau$ tested: the cycled-aged cross-section consistently shows the highest late intergranular area fraction, width, and tortuosity, while calendar-aged late-crack coverage remains comparable to the initial state throughout. Within $\tau\in[0.05, 0.2]$ the absolute morphometric values closely track the main-text numbers; outside this range the relative ordering is preserved but absolute values deviate more substantially. The operating point $\tau{=}0.1$ used throughout this study is highlighted.}
  \label[extfigure]{fig:ext_tau_sensitivity}
\end{figure}

\begin{figure}[htbp]
  \centering
  \includegraphics[width=0.9\linewidth]{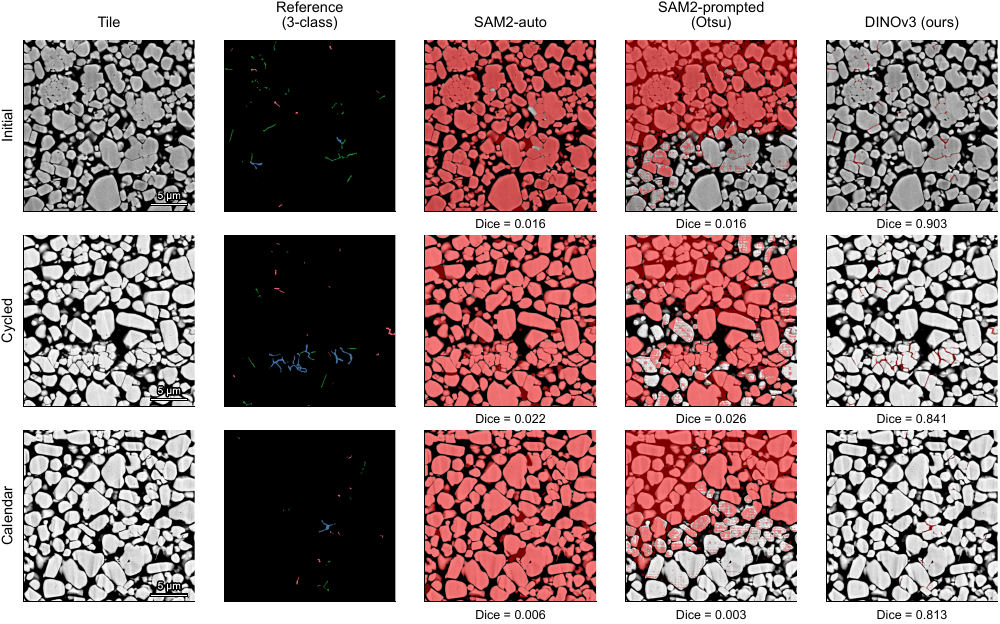}
  \caption{\textbf{$|$ Zero-shot SAM2 vs.\ fine-tuned DINOv3 on the NMC crack validation set.} Columns: raw tile; expert 3-class reference (intragranular red, early intergranular green, late intergranular blue); SAM2 in automatic mode ($64\times64$ point grid, no prompts); SAM2 prompted with Otsu-derived dark-region centroids and area-bounded lowest-mean-intensity multi-mask selection; fine-tuned DINOv3-DeepLabV3+. Per-prediction binary Dice values are annotated below each prediction column. The inset on the SAM2-auto column of row~1 shows the unfiltered output before geometric and intensity post-processing, illustrating the oversegmentation failure mode on textured SEM matrix. The 3-class reference is collapsed to a single crack vs.\ background target because SAM2 is class-agnostic; this typology gap is the structural reason fine-tuning is required for the cycled/initial late-intergranular comparison.}
  \label[extfigure]{fig:ext_sam_baseline}
\end{figure}

\begin{figure}[htbp]
  \centering
  \includegraphics[width=0.9\linewidth]{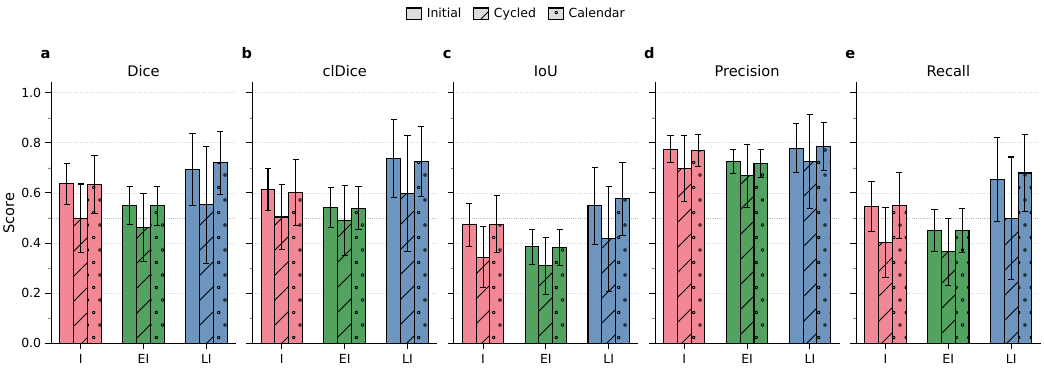}
  \caption{\textbf{$|$ Per-cross-section segmentation metrics by aging condition.} Mean per-class scores for Dice, clDice, IoU, Precision, and Recall (with standard-deviation error bars) computed separately for the initial, cycled-aged, and calendar-aged cross-sections. Crack classes are abbreviated as I (intragranular), EI (early intergranular), and LI (late intergranular). The relative ranking of per-class scores is preserved across aging conditions: the frozen DINOv3 encoder transfers across the three aging states despite substantial differences in SEM contrast, brightness, and late-crack density.}
  \label[extfigure]{fig:ext_cross_sample}
\end{figure}

\begin{table}[htbp]
\caption{\textbf{$|$ Zero-shot SAM2 baselines vs.\ fine-tuned DINOv3 (binary segmentation).} SAM2 \texttt{sam2.1-hiera-large} in two configurations evaluated on the same 482-tile reference set as the main DINOv3 result. Binary Dice and IoU were computed after collapsing the 3-class reference to crack-vs-background. For each SAM2 configuration the higher of the with- and without-filter values was retained per tile. The 95\% CIs are non-paired bootstrap with $n=1000$ resamples.}\label{tab:extsambaseline}
\resizebox{\textwidth}{!}{%
\begin{tabular}{lccc}
\toprule
Method & Binary Dice $\uparrow$ & Binary IoU $\uparrow$ & 3-class capable? \\
\midrule
DINOv3-DeepLabV3+ (ours, binary) & 0.77 & 0.65 & \checkmark\,(via 3-class output) \\
SAM2-auto (zero-shot) & 0.019 & 0.010 & --- \\
SAM2-prompted (Otsu) & 0.025 & 0.013 & --- \\
\bottomrule
\end{tabular}}
\end{table}

\begin{table}[htbp]
\caption{\textbf{$|$ Morphometric statistics by sample and crack class.} Mean $\pm$ std across ROIs, supporting the per-condition morphometric comparisons in the main article. Bold: highest values within each metric column. Pixel size: 22.3~nm.}\label{tab:extmorphometrics}
\footnotesize
\setlength{\tabcolsep}{3pt}
\resizebox{\textwidth}{!}{%
\begin{tabular}{llccccc}
\toprule
\textbf{Sample} & \textbf{Class} & \textbf{ROIs} & \textbf{Area (\%)} & \textbf{Width (nm)} & \textbf{Length ($\mu$m)} & \textbf{Tortuosity} \\
\midrule
Initial & Intragranular & 446 & $10.34 \pm 7.54$ & $107 \pm 33$ & $0.8 \pm 0.6$ & $0.95 \pm 0.31$ \\
Initial & Intergran.\ early & 97 & $1.66 \pm 0.76$ & $91 \pm 10$ & $8.6 \pm 5.7$ & $1.38 \pm 0.53$ \\
Initial & Intergran.\ late & 97 & $0.52 \pm 1.29$ & $159 \pm 22$ & $1.5 \pm 3.4$ & $1.55 \pm 0.63$ \\
\midrule
Cycled-aged & Intragranular & 610 & $9.25 \pm 7.17$ & $107 \pm 32$ & $1.1 \pm 0.9$ & $0.97 \pm 0.31$ \\
Cycled-aged & Intergran.\ early & 69 & $1.30 \pm 1.23$ & $98 \pm 15$ & $7.3 \pm 7.4$ & $1.58 \pm 0.73$ \\
Cycled-aged & Intergran.\ late & 69 & $\mathbf{4.58 \pm 3.60}$ & $\mathbf{167 \pm 39}$ & $\mathbf{14.2 \pm 10.5}$ & $\mathbf{2.40 \pm 1.00}$ \\
\midrule
Calendar-aged & Intragranular & 623 & $9.51 \pm 7.96$ & $106 \pm 35$ & $0.9 \pm 0.7$ & $0.93 \pm 0.32$ \\
Calendar-aged & Intergran.\ early & 49 & $2.34 \pm 1.01$ & $99 \pm 11$ & $10.2 \pm 5.1$ & $1.71 \pm 0.47$ \\
Calendar-aged & Intergran.\ late & 49 & $0.49 \pm 1.16$ & $146 \pm 26$ & $1.4 \pm 3.3$ & $1.45 \pm 0.63$ \\
\bottomrule
\end{tabular}}
\end{table}

\clearpage
\section{Supplementary information}
\label{app:si}

\emph{This appendix reproduces the Supplementary Information accompanying the
manuscript. References to ``the main article'' below refer to
Sections~\ref{sec:main}--\ref{Methods} of this preprint; Extended Data items are
in \hyperref[app:ed]{Appendix~A}.}

\setcounter{figure}{0}
\setcounter{table}{0}
\renewcommand{\thefigure}{S\arabic{figure}}
\renewcommand{\thetable}{S\arabic{table}}
\renewcommand{\theHfigure}{SI.\arabic{figure}}
\renewcommand{\theHtable}{SI.\arabic{table}}
\renewcommand{\figurename}{Supplementary Figure}
\renewcommand{\tablename}{Supplementary Table}
\crefname{figure}{Supplementary Figure}{Supplementary Figures}
\Crefname{figure}{Supplementary Figure}{Supplementary Figures}
\crefname{table}{Supplementary Table}{Supplementary Tables}
\Crefname{table}{Supplementary Table}{Supplementary Tables}

\section*{Supplementary Notes}

\subsection*{Cell, aging, and post-mortem sample preparation}\label{si:cell-prep}

Aged samples were prepared from commercial Hunan LiFun Technology pouch cells (Model 103962) with a Ni-rich NCM cathode and graphite anode. Per the manufacturer datasheet (PS-FC-103962-01, rev.~A0, 2018-11-05), each cell has nominal capacity 3400\,mAh, nominal voltage 3.80\,V, operating window 3.0--4.35\,V, internal impedance $\leq$60\,m$\Omega$ (50\,\% SoC, 1\,kHz~AC), and Al-laminate pouch dimensions $\leq$10.4\,$\times$\,39.0\,$\times$\,62.0\,mm at $\sim$50\,g. The electrolyte is a LiPF$_6$ solution in EC/EMC/DMC, and the separator is a polyethylene film with a ceramic coating; the cells are qualified to GB/T 18287-2013 and GB 31241-2014.

Three cells were sampled to span representative aging states: an initial (formation-only) reference cell, a cycled-aged cell (K100, 100 charge--discharge cycles), and a calendar-aged cell (K187, stored at $45^\circ$C). One cell per condition was prepared for post-mortem cross-sectioning. Capacity retention at the end of aging was measured by reference performance tests (RPT) at C/3 ($\approx$1.13\,A relative to the 3.4\,Ah nominal capacity): the two characterised initial cells delivered 3.59 and 3.68\,Ah (mean 3.64\,Ah); K100 retained 3.12\,Ah (86\,\% of initial mean); K187 retained 2.68\,Ah (74\,\% of initial mean). After aging and the final RPT, cells were transferred to an argon-filled glovebox (O$_2$ and H$_2$O $<$0.5\,ppm), disassembled, and the cathode coupons cross-sectioned and mounted for SEM imaging.

\subsection*{Cross-section SEM imaging}\label{si:sem}

Cross-section images of the three NMC cathodes were acquired on a ZEISS Supra~55 field-emission scanning electron microscope operated at 10\,kV accelerating voltage using the backscattered electron (BSE) detector. Image tiles were assembled with ZEISS Atlas~5 (v5.3.5) with no line or frame averaging (LineAvg = 1). Pixel size was nominally 22.3\,nm/px (per-sample: 22.29\,nm/px for K100, 22.38\,nm/px for K187, 22.28\,nm/px for the initial sample). Final image dimensions were $17{,}088 \times 6{,}595$\,px (K100), $17{,}017 \times 6{,}848$\,px (K187), and $17{,}137 \times 7{,}015$\,px (initial). Acquisition parameters were identical across the three samples.

\subsection*{S1.\,Training configuration}
The encoder was kept fixed; only the decoder and output head were optimised during training. We used the AdamW optimiser~\cite{loshchilov2019adamw} with learning rate $\eta{=}10^{-4}$, $\beta_1{=}0.9$, $\beta_2{=}0.98$, $\varepsilon{=}10^{-9}$, and weight decay $\lambda{=}10^{-3}$. A cosine-cyclical learning-rate schedule (4 warm-restart cycles, minimum learning rate~$5{\times}10^{-5}$) was applied over the full training run. Training ran for up to 500 epochs with early stopping (patience~$=$~20 epochs, monitoring validation loss). The batch size was 4 at $1024{\times}1024$~px in single precision (FP32), with a fixed random seed of 42. Model selection retained the checkpoint with the lowest validation loss; the selected checkpoint corresponds to epoch~465. Online data augmentation was applied during training to improve robustness to SEM acquisition variation while preserving crack topology; it comprised random affine transforms, flips, crop-and-resize, brightness/contrast jitter, additive Gaussian noise and blur, salt-and-pepper noise, pixel dropout, and grid masking, with full per-operation probabilities and ranges listed in Supplementary Section~S8.

We minimised a per-channel weighted sum of binary cross-entropy (BCE), mean squared error (MSE), and Dice losses. Let $y_c(x)\in\{0,1\}$ denote the target mask and $p_c(x)\in[0,1]$ the predicted probability for class $c$ at pixel $x$, with class weights $w_c$ and $\varepsilon>0$ for numerical stability:
\begin{align}
\mathcal{L}_{\text{BCE}} &= -\frac{1}{|\Omega|}\sum_{x\in\Omega}\sum_{c=1}^{C} w_c\Big[y_c(x)\log p_c(x) + \big(1-y_c(x)\big)\log\big(1-p_c(x)\big)\Big],\\
\mathcal{L}_{\text{MSE}} &= \frac{1}{|\Omega|}\sum_{x\in\Omega}\sum_{c=1}^{C} w_c\big(p_c(x)-y_c(x)\big)^2,\\
\mathcal{L}_{\text{Dice}} &= 1 - \frac{1}{\sum_{c=1}^{C} w_c}\sum_{c=1}^{C} w_c\,
\frac{2\sum_{x\in\Omega} p_c(x)\,y_c(x)+\varepsilon}{\sum_{x\in\Omega} p_c(x) + \sum_{x\in\Omega} y_c(x) + \varepsilon},
\end{align}
with the total objective
\begin{equation}
\mathcal{L} \;=\; \lambda_{\text{BCE}}\,\mathcal{L}_{\text{BCE}} \;+\; \lambda_{\text{MSE}}\,\mathcal{L}_{\text{MSE}} \;+\; \lambda_{\text{Dice}}\,\mathcal{L}_{\text{Dice}},
\end{equation}
where $\lambda_{\text{BCE}}=\lambda_{\text{MSE}}=\lambda_{\text{Dice}}=1$. Equal weights were used as a neutral initialisation, treating all three loss components as equally informative a priori. This symmetric combination is consistent with multi-term loss formulations commonly adopted in biomedical segmentation~\cite{ronneberger2015unet,deeplabv3plus_arxiv}.

\subsection*{S2.\,Image-scale inference, stitching, and ROI extraction}
We applied the trained model to full-resolution SEM images to generate per-pixel class-probability maps and corresponding overlay masks. Each image ($\sim 17{,}000 \times 7{,}000$~px) was read in greyscale, rescaled to $[0,1]$, and processed in evaluation mode using the frozen DINOv3 encoder~\cite{dinov3_arxiv} with the trained DeepLabV3+ decoder~\cite{deeplabv3plus_arxiv}. Because the field of view exceeds GPU memory, inference was performed with an overlapping sliding-window strategy (50\% overlap). Per-tile predictions $p^{(t)}_c(x)$ were stitched by visit-normalised averaging,
\begin{equation}
\bar{p}_c(x) \;=\; \frac{1}{N(x)}\sum_{t\in\mathcal{T}} \mathbf{1}\{x\in t\}\; p^{(t)}_c(x),
\quad N(x)=\sum_{t\in\mathcal{T}} \mathbf{1}\{x\in t\},
\label{eq:stitch}
\end{equation}
which suppresses artefacts at tile boundaries. To reduce false responses on bright foreground regions, we optionally gated the stitched maps using the greyscale intensity $g(x)$ and a background threshold $\tau_g$, chosen either as a fixed value or from Otsu's method $\tau_g=\kappa\,\tau_{\text{Otsu}}$~\cite{otsu1979}. Hard gating sets predictions outside the dark background to zero,
\begin{equation}
\tilde{p}_c(x)\;=\;\bar{p}_c(x)\,\mathbf{1}\{g(x)\le\tau_g\},
\label{eq:hardgate}
\end{equation}
whereas soft gating down-weights them smoothly:
\begin{equation}
\tilde{p}_c(x)\;=\;\bar{p}_c(x)\;\mathrm{clip}\!\left(\frac{\tau_g-g(x)}{\max(\tau_g, \epsilon)},\, 0,\, 1\right).
\label{eq:softgate}
\end{equation}
A light Gaussian blur on $g$ and brief morphological cleanup of the gate mask reduced artefacts without eroding thin cracks. Final per-class binary masks were obtained by global thresholding, $M_c(x)=\mathbf{1}\{\tilde{p}_c(x)\ge\tau\}$, and exported as per-class masks and composite overlays. Overlays were rendered by alpha blending at native resolution for side-by-side comparison across the \emph{initial}, \emph{calendar-aged}, and \emph{cycled-aged} probes.

Following stitching, connected components in the binary maps are filtered by a minimum pixel area and a minimum skeleton length, suppressing isolated artefacts while preserving thin, elongated cracks. Particle-level ROIs are then extracted by intensity-thresholding the SEM backscatter contrast (cathode brighter than epoxy matrix), followed by morphological closing, area filtering, and padding, with each ROI clipped to an in-bounds square window. ROIs are ordered top-to-bottom and left-to-right for reproducible enumeration across cross-sections (panels a--e of \cref{fig:si_roi_detection}).

\subsection*{S3.\,Segmentation operating point and threshold selection}
All quantitative results in the main article use the decision threshold $\tau{=}0.1$. This operating point was selected on the validation precision--recall (PR) curve to maximise the class-averaged Dice score under the highly imbalanced foreground fractions characteristic of thin crack structures in SEM cross-sections (crack pixels constitute less than 1\% of the tile area). At higher thresholds, recall collapses faster than the false-positive rate is reduced, so the class-averaged Dice peaks toward the low end of the threshold sweep. The PR-based threshold selection therefore yields a recall-sensitive operating point appropriate for sparse foreground structures; the corresponding PR curves and threshold sweep are shown in \cref{fig:ext_pr_threshold} of the main article, and the resulting robustness of morphometric outputs to the choice of $\tau$ in the interval $[0.05, 0.2]$ is documented in \cref{fig:ext_tau_sensitivity} of the main article. Applying the optional intensity-based background gating (Eqs.~\ref{eq:hardgate}--\ref{eq:softgate}) at this operating point raises precision to 0.877 at the cost of recall (0.302), lowering the mean Dice to 0.434 and reflecting the false-positive/recall trade-off on bright cathode regions.

\subsection*{S4.\,Supplementary morphometric definitions}
The main Methods define the principal morphometric descriptors used in the analysis (crack length, area fraction, orientation/anisotropy, geometric tortuosity, and transverse roughness). For completeness, we record here the standard-operation conventions adopted by the pipeline.

\begin{figure}[!htbp]
  \centering
  \includegraphics[width=\linewidth]{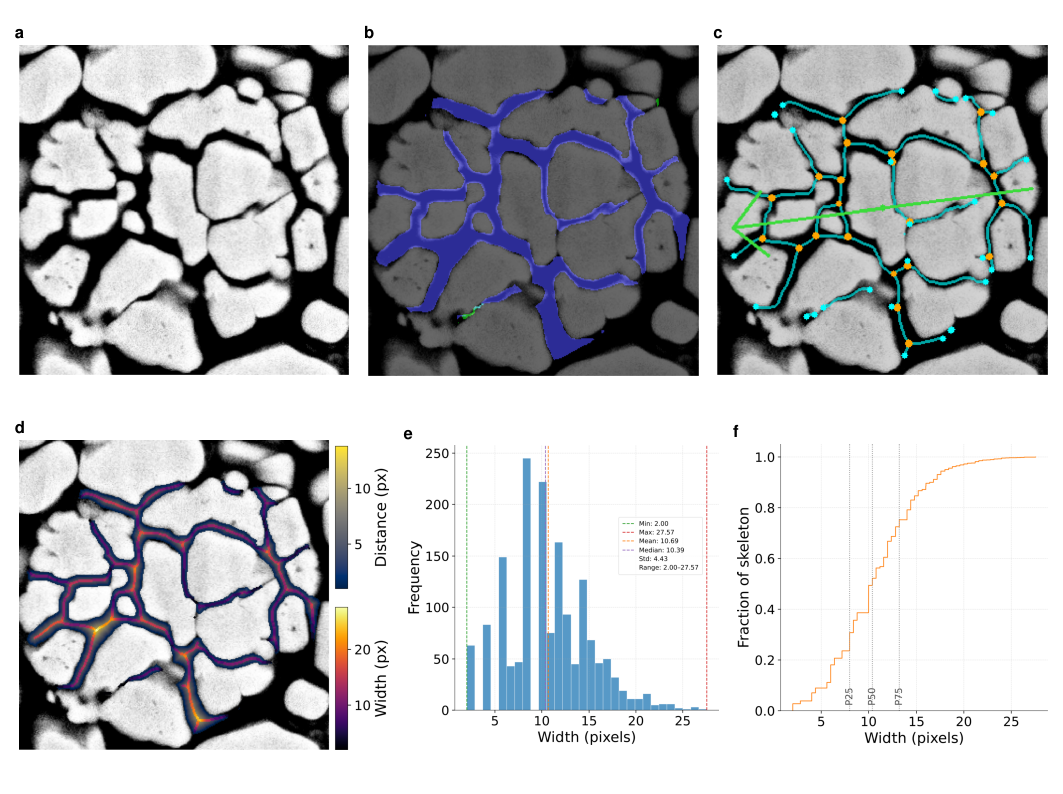}
  \caption{\textbf{Skeleton-based morphometric extraction.} The skeleton-based extraction procedure applied within each ROI, illustrated for an example \emph{intergranular late} ROI. (a)~Raw SEM image of the region of interest. (b)~Multi-class segmentation overlay fused with a semi-transparent highlight of the single-class crack mask (white: 16.1\% of pixels, 17\,666~px; black: 83.9\%, 91\,895~px). (c)~Medial-axis skeleton overlaid on the raw SEM image, with branch points (orange) and endpoints (cyan) classified from the $3\times3$ neighbour-degree topology and the principal orientation (172.6\textdegree) derived from the leading eigenvector of the skeleton coordinate covariance matrix; the network comprises 74 branches and 30 endpoints, with tortuosity 5.39, anisotropy 1.51, and roughness 64.76~px. (d)~Fused distance-transform map (cividis colour scale) and per-skeleton-pixel width heatmap $w(x_s)=2\,d(x_s)$ (inferno colour scale), both overlaid on the raw ROI. (e)~Crack width histogram (mean 10.69~px, median 10.39~px, std 4.43~px, range 2.00--27.57~px). (f)~Empirical cumulative distribution function of crack width with $P_{25}$, $P_{50}$, and $P_{75}$ percentiles indicated. Together, panels (c)--(f) document that the pipeline reliably extracts centrelines even for branched, tortuous crack networks and yields stable width estimates along elongated crack segments.}
  \label{fig:si_skeleton}
\end{figure}

Skeleton-pixel topology was classified by the local neighbourhood degree on the eight-connected $3\times3$ window centred on each foreground skeleton pixel. Pixels with degree~1 were labelled \emph{endpoints} (crack termini), pixels with degree~$\ge3$ were labelled \emph{branch points} (crack junctions), and pixels with degree~0 were labelled \emph{isolated} (and excluded from path-length statistics). This convention is the standard medial-axis classification used throughout connected-component morphometry and corresponds to the topology counts reported in \cref{fig:si_skeleton}.

Width statistics derived from the Euclidean distance transform are reported using the empirical cumulative distribution function (ECDF)
\begin{equation*}
\widehat{F}(t)\;=\;\frac{1}{N}\sum_{i=1}^{N}\mathbf{1}\{w_i \le t\},
\end{equation*}
i.e.\ the fraction of the centreline length with width $\le t$~\cite{Wasserman2004}. We summarise each ECDF by the first, second, and third quartiles ($P_{25}$, $P_{50}$, $P_{75}$); these robust descriptors are reported in preference to the mean and standard deviation for cross-ROI comparisons because the underlying width distributions are typically heavy-tailed and right-skewed by junction pixels.

\subsection*{S5.\,Spatial overlap quantification}
Validation tiles were quantified for spatial overlap with training tiles. Each validation tile $v$ was assigned an overlap fraction
\begin{equation*}
o(v) \;=\; \max_{t} \, \frac{|v \cap t|}{|v|}
\end{equation*}
taken over all training tiles $t$ from the same SEM cross-section. By construction of the 50\%-overlap sliding-window tiling, adjacent tiles share 50\% of their area. Under the random 80/20 per-image partition, the probability that all four cardinal neighbours of a validation tile lie in the validation set is $0.2^4 \approx 0.002$, so the median overlap fraction $o(v)$ across the 96 validation tiles is $\approx 0.50$ by construction. The reported pixel-level segmentation metrics are therefore most meaningfully read as operating-point calibration on the full 482-tile expert-corrected reference set rather than as out-of-distribution generalisation error; the morphometric distributions reported in the main article are computed at the ROI level on the full cross-sections and are independent of the tile-level split.

Image-disjoint splits were not used in this initial study because only three SEM cross-sections were available (one per aging condition), making a leave-one-condition-out split infeasible without losing condition coverage during training. Removing any single image from the training pool would deprive the model of one of the three target aging states and would prevent the validation set from sampling that condition altogether; the corresponding metric estimates would therefore conflate cross-condition generalisation error with the loss of in-distribution training signal. A rotating image-disjoint sensitivity analysis, in which the model is retrained with each cross-section held out in turn and the remaining two used for both training and within-image validation, would require additional independent cross-sections per condition and is left for future work as more high-resolution SEM data become available.

\subsection*{S6.\,SAM2 baseline details}\label{si:sam2-details}
The SAM2 (Hiera-Large, \texttt{sam2.1}) baseline was evaluated in two configurations. In automatic mode, a $64\times64$ dense point grid was used with no other prompts, and the union of SAM2's returned masks was taken as the raw output. In prompted mode, point prompts were placed at the centroids of dark regions identified by Otsu thresholding ($\text{otsu\_factor}=1.5$), with up to 128 prompts per tile. For each prompted point, the area-bounded candidate with the lowest mean intensity (candidate area $\leq 0.25$ of the tile) was selected from SAM2's three multi-mask outputs. Geometric and intensity post-filters (mask elongation $\geq 2.0$, solidity $\leq 0.85$, mean intensity $\leq 0.45$) were applied per tile in both configurations, and the higher of the with- and without-filter Dice was retained per tile.

The comparison was performed on a binary crack-versus-background target obtained by collapsing the three-class reference, mirroring SAM2's class-agnostic output. A heuristic post-hoc width-based class assignment was considered and rejected, since it would conflate SAM2's segmentation quality with the supervised classifier reported in the main article.

\subsection*{S7.\,Statistical-analysis bookkeeping}\label{si:stats-bookkeeping}
The family of tests reported in the main article comprises 27 comparisons: 3 ROI subsets (intergranular, intragranular, all cracks) $\times$ 3 morphometric metrics (area fraction, mean width, geometric tortuosity) $\times$ 3 condition pairs (initial vs cycled-aged, initial vs calendar-aged, cycled-aged vs calendar-aged). Benjamini--Hochberg-adjusted $q$-values within this 27-test family appear as significance brackets in the morphometric figure of the main article (ROI counts: $194/138/98$ for intergranular early $+$ late, $446/610/623$ for intragranular, initial / cycled-aged / calendar-aged).

A separate 9-test sub-family is reported on the intergranular-late subset ($97/69/49$ ROIs), again Benjamini--Hochberg-adjusted within that sub-family. Both families are reported because the headline cycling-versus-baseline claim is tested under each scope and survives FDR adjustment at $q<0.001$ in both.

Raw $U$-statistics, raw $p$-values, and adjusted $q$-values for all 27 comparisons are listed in the analysis bundle under \texttt{fig5\_statistics\_pairwise\_stats.tsv} in the released code repository.

\subsection*{S8.\,Software environment and reproducibility}\label{si:reproducibility}
All models were trained and evaluated on a single NVIDIA A100 80\,GB PCIe GPU (driver 550.163.01) under Ubuntu 22.04.5 LTS, using Python 3.10.13, PyTorch 2.4.0 (CUDA 12.1, cuDNN 9.1.0), torchvision 0.19.0, PyTorch-Lightning 2.0.8, timm 0.9.6, scikit-image 0.25.2, SciPy 1.15.3, NumPy 1.26.4, and OpenCV 4.13. The frozen DINOv3 ViT-B/16 encoder fed a DeepLabV3+ decoder (ASPP dilation rates 6/12/18, low-level stride 2), giving 17.8\,M trainable and 85.7\,M frozen parameters.

The optimiser, learning-rate schedule, batch size, precision, and loss formulation are specified in Supplementary Section~S1 (the Dice term used a smoothing constant $\varepsilon{=}1$). Online data augmentation was applied to the training split only; validation data were left unaltered. The training framework's default pipeline was applied to every training sample, with each operation drawn independently at its own probability: random affine transforms (probability~0.5; rotation up to $\pm45^\circ$, translation up to half the image extent, isotropic scaling in $[0.9,1.1]$); random crop-and-resize (0.1; crop area fraction $[0.6,0.95]$); horizontal and vertical flips (0.5 each); brightness and contrast jitter (0.1 each; factor $[0.9,1.1]$); additive Gaussian noise (0.1; per-call mean in $[0,0.1]$ and standard deviation in $[0,0.2]$, clipped to $[0,1]$); Gaussian blur (0.1; odd kernel size 3--7\,px, $\sigma\in[0.1,2.0]$); salt-and-pepper noise (0.1; salt and pepper fractions each in $[0,0.01]$); random pixel dropout (0.1; fraction in $[0,0.05]$); and grid masking (0.1; up to 10 zeroed cells on a $10\times10$ grid). Geometric operations (affine, crop, flips) were applied jointly to the image and mask, and photometric operations to the image only; hue and saturation jitter present in the default configuration are inactive for the single-channel (greyscale) SEM input used here.

Inference and ROI extraction followed Supplementary Sections~S2--S3 with the following operating values. Per-class probability maps were binarised at $\tau{=}0.1$. Particle ROIs were obtained by Otsu thresholding of the backscatter image after a $5\times5$ Gaussian pre-blur (no scaling factor), morphological closing with a $5\times5$ elliptical kernel, and a 50\,px fragment-merge radius, retaining particle components with area $\geq$10{,}000\,px and crack components with area $\geq$1000\,px and skeleton length $\geq$100\,px, with zero ROI padding. Centrelines were extracted by Zhang--Suen thinning (\texttt{skimage.morphology.skeletonize}) and widths from the L2 distance transform as $w=2\,d$. The intensity-based background gating of Eqs.~\ref{eq:hardgate}--\ref{eq:softgate} is an optional step that was disabled in the production pipeline. All code and trained weights are available as stated in the Code Availability section.

\clearpage
\section*{Supplementary Figures}

\clearpage
\begin{figure}[htbp]
  \centering
  \includegraphics[width=\linewidth]{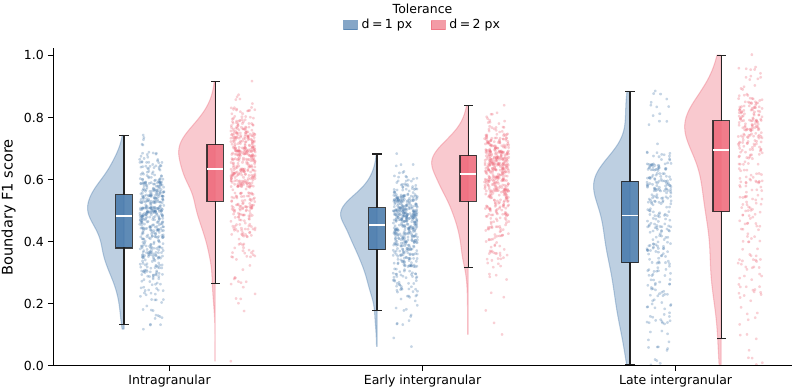}
  \caption{\textbf{Per-tile boundary-F1 distribution at 1\,px and 2\,px tolerance.} Per-tile boundary-F1 (BF1) score distributions at 1-pixel and 2-pixel Euclidean boundary tolerance for each crack class. The median BF1 at 1-pixel tolerance (0.45--0.46 per class) rises to 0.58--0.62 at 2-pixel tolerance, approaching the clDice level. The improvement from 1\,px to 2\,px confirms that 1--2-pixel boundary misalignment dominates the moderate pixel-exact Dice. For crack structures only 2--5 pixels wide at the imaging resolution of 22.3~nm/px, this boundary uncertainty corresponds to sub-50~nm localisation error, comparable to the precision achievable by human annotators tracing approximate crack centrelines.}
  \label{fig:si_bf1_distribution}
\end{figure}

\clearpage
\begin{figure}[htbp]
  \centering
  \includegraphics[width=\linewidth]{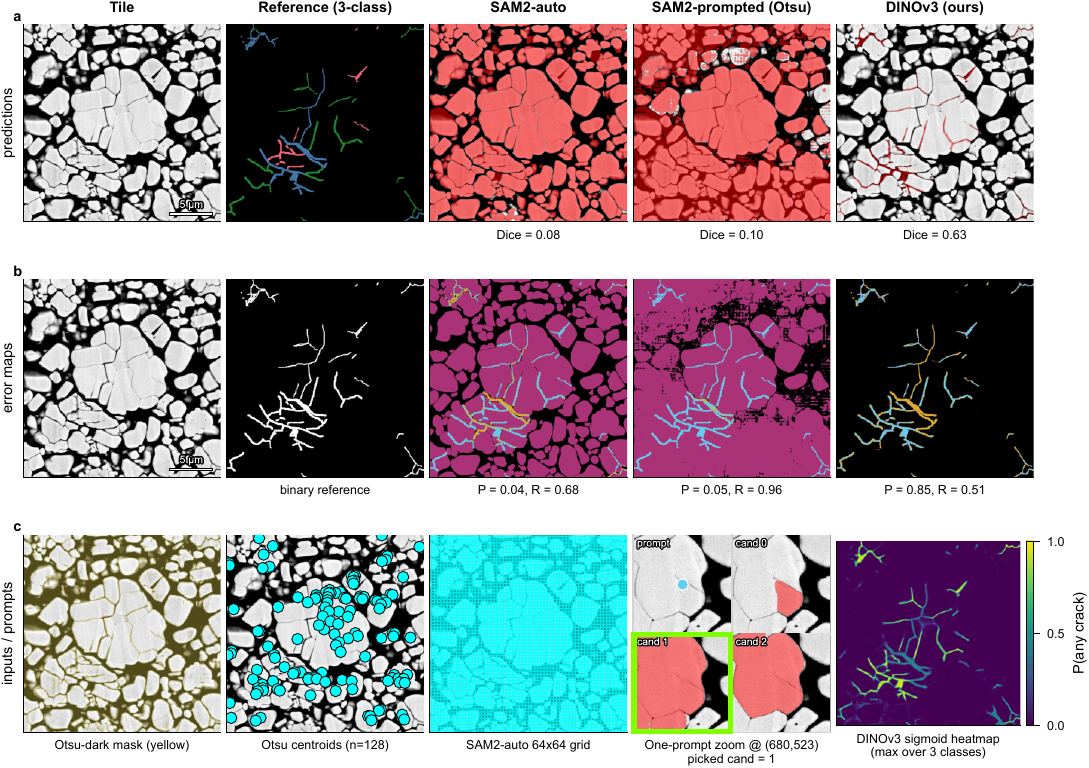}
  \caption{\textbf{Method-by-method diagnostic on tile \texttt{100\_0000\_00038} (the Fig.~1 sample tile).} \textbf{Row a (predictions):} columns left-to-right are the raw tile, the 3-class expert reference (intragranular red, early intergranular green, late intergranular blue), SAM2-auto, SAM2-prompted, and fine-tuned DINOv3; each prediction cell is annotated with its binary Dice. \textbf{Row b (error maps):} same column order; cyan = true positive, magenta = false positive, amber = false negative, black = true negative. \textbf{Row c (inputs):} for each SAM2 column, the prompts the method sees: Otsu-derived dark mask overlay (column 1), Otsu centroids (column 2, SAM2-prompted inputs), the $64\times 64$ dense point grid (column 3, SAM2-auto inputs), and one prompt zoom showing SAM2's 3 multi-mask candidates with the area-bounded lowest-mean-intensity selection bordered in green (column 4); column 5 shows the DINOv3 sigmoid heatmap (maximum over the 3 crack classes) before binarisation, with the colour bar indicating $P(\textnormal{any crack})$ per pixel.}
  \label{fig:si-sam-baseline-zoom}
\end{figure}

\clearpage
\begin{figure}[htbp]
  \centering
  \includegraphics[width=\linewidth]{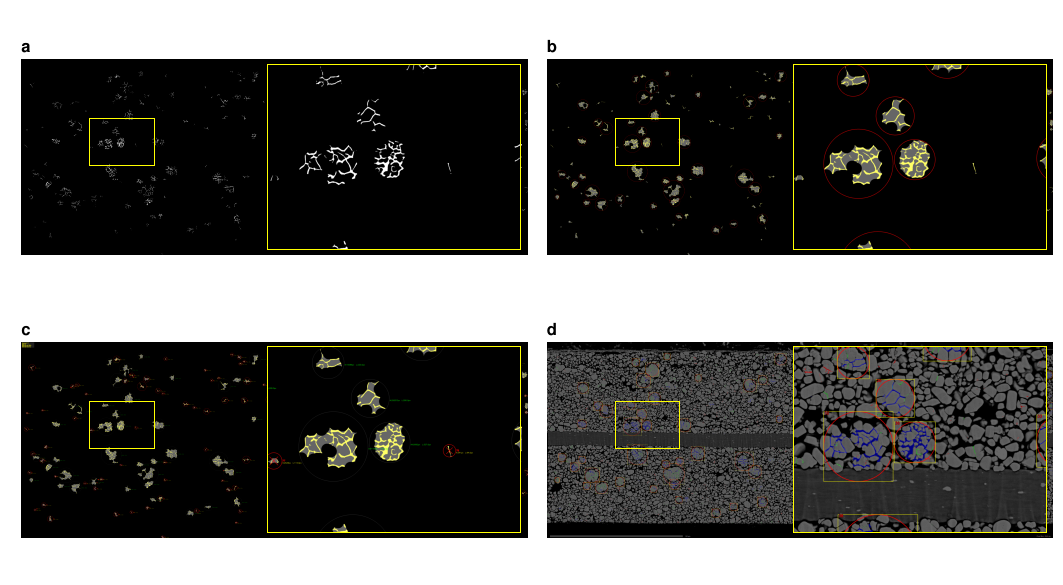}
  \caption{\textbf{Region-of-interest detection pipeline.} The automated ROI-extraction workflow that defines the spatial units used for the morphometric analysis in the main article, illustrated for the cycled-aged image (intergranular late class). (a)~Binary threshold mask at decision threshold $\tau$. (b)~Post-merge mask after morphological closing merges spatially proximal crack fragments belonging to the same secondary particle. (c)~Rejected sub-particle components below the minimum-area threshold, suppressed to remove artefacts. (d)~Final detected ROI windows: each surviving component is enclosed by its axis-aligned bounding box, expanded by a fixed margin to ensure full particle coverage. The resulting ROI windows are sorted in raster order (top-to-bottom, left-to-right) and used as the spatial units for all downstream morphometric statistics reported in the main article.}
  \label{fig:si_roi_detection}
\end{figure}

\clearpage
\begin{figure}[htbp]
  \centering
  \includegraphics[width=\linewidth]{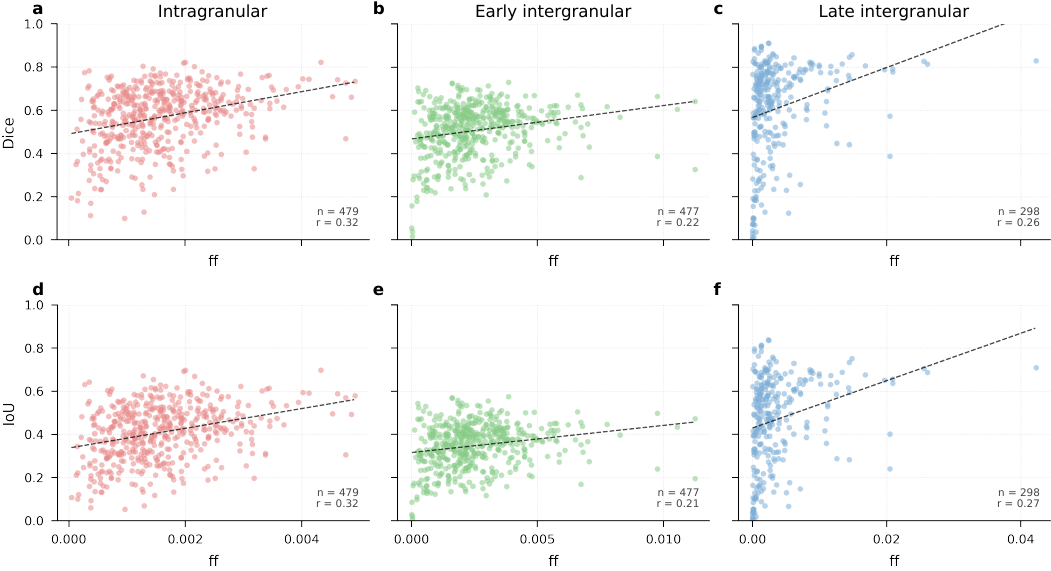}
  \caption{\textbf{Per-class segmentation metrics versus crack density (full panel set).} Per-tile Dice (top row) and IoU (bottom row) plotted against the foreground pixel fraction for intragranular (\textbf{a},\,\textbf{d}), intergranular early (\textbf{b},\,\textbf{e}), and intergranular late (\textbf{c},\,\textbf{f}) cracks across all 482 expert-corrected tiles. Dashed lines show linear fits; Pearson correlation coefficients $r$ are annotated in each panel ($r=0.21$--$0.32$). The full multi-panel decomposition shown here supports the conclusion that pixel-exact agreement is principally limited by low-density tiles rather than by model failures on crack-rich regions, motivating the topology-aware and boundary-tolerant metrics (clDice, BF1) reported in the main article.}
  \label{fig:si_metric_vs_density}
\end{figure}

\newpage
\bibliographystyle{elsarticle-num}
\bibliography{references}

\end{document}